\documentclass[aps,pre,twocolumn,groupedaddress,superscriptaddress,showpacs]{revtex4-1}

\usepackage{graphicx}
\usepackage{color}
\usepackage{amsmath}
\usepackage{amsfonts}
\usepackage{amssymb}
\usepackage{dcolumn}
\usepackage{url,lineno}

\begin{document}

\title{Watersheds in disordered media}

\author{N. A. M. Ara\'ujo}
\email{nmaraujo@fc.ul.pt}
\affiliation{Centro de F\'isica Te\'orica e Computacional, Departamento
de F\'isica, Faculdade de Ci\^encias, Universidade de Lisboa, Lisboa,
Portugal} 

\author{K. J. Schrenk}
\email{kjs73@cam.ac.uk}
\affiliation{Department of Chemistry, University of Cambridge,
Cambridge, United Kingdom}

\author{H. J. Herrmann}
\email{hans@ifb.baug.ethz.ch}
\affiliation{Computational Physics, Institute for Building Materials, ETH Zurich, Zurich, Switzerland}
\affiliation{Departamento de F\'isica, Universidade Federal do Cear\'a,
Fortaleza, Cear\'a, Brazil}

\author{J. S. Andrade Jr}
\email{soares@fisica.ufc.br}
\affiliation{Departamento de F\'isica, Universidade Federal do Cear\'a,
Fortaleza, Cear\'a, Brazil}
\affiliation{Computational Physics, Institute for Building Materials, ETH Zurich, Zurich, Switzerland}

\begin{abstract}
What is the best way to divide a rugged landscape? Since ancient times,
watersheds separating adjacent water systems that flow, for example, toward
different seas, have been used to delimit boundaries.  Interestingly, serious
and even tense border disputes between countries have relied on the subtle
geometrical properties of these tortuous lines. For instance, slight and even
anthropogenic modifications of landscapes can produce large changes in a
watershed, and the effects can be highly nonlocal. Although the watershed
concept arises naturally in geomorphology, where it plays a fundamental role in
water management, landslide, and flood prevention, it also has important
applications in seemingly unrelated fields such as image processing and
medicine.  Despite the far-reaching consequences of the scaling properties on
watershed-related hydrological and political issues, it was only recently that
a more profound and revealing connection has been disclosed between the concept
of watershed and statistical physics of disordered systems. This review
initially surveys the origin and definition of a watershed line in a
geomorphological framework to subsequently introduce its basic geometrical and
physical properties. Results on statistical properties of watersheds
obtained from artificial model landscapes generated with long-range
correlations are presented and shown to be in good qualitative and quantitative
agreement with real landscapes.
\end{abstract}

\maketitle

\section{Introduction}\label{sec::intro}

Although both start in the mountains of Switzerland, the Rhine
and Rhone rivers diverge while flowing towards different seas. While
the Rhine empties into the North Sea, the Rhone drains into the
Mediterranean. Similarly, all over the world one finds rivers with close
sources but distant mouths. For example, the Colorado and the Rio Grande
even open towards different oceans. When looking at a landscape,
how to identify the regions draining towards one side or the other? When
rain falls or snow melts on a landscape, the dynamics of the surface
water is determined by the topography of the landscape.  Water flows
downhill and overpasses small mounds by forming lakes which eventually
overspill. A drainage basin is then defined as the region where water
flows towards the same outlet. Its shape and extension strongly depend
on the topography.  The line separating two adjacent basins is the
watershed line, which typically wanders along the mountain
crests~\cite{Gregory73}.

\begin{figure}
\begin{center}
\includegraphics[width=\columnwidth]{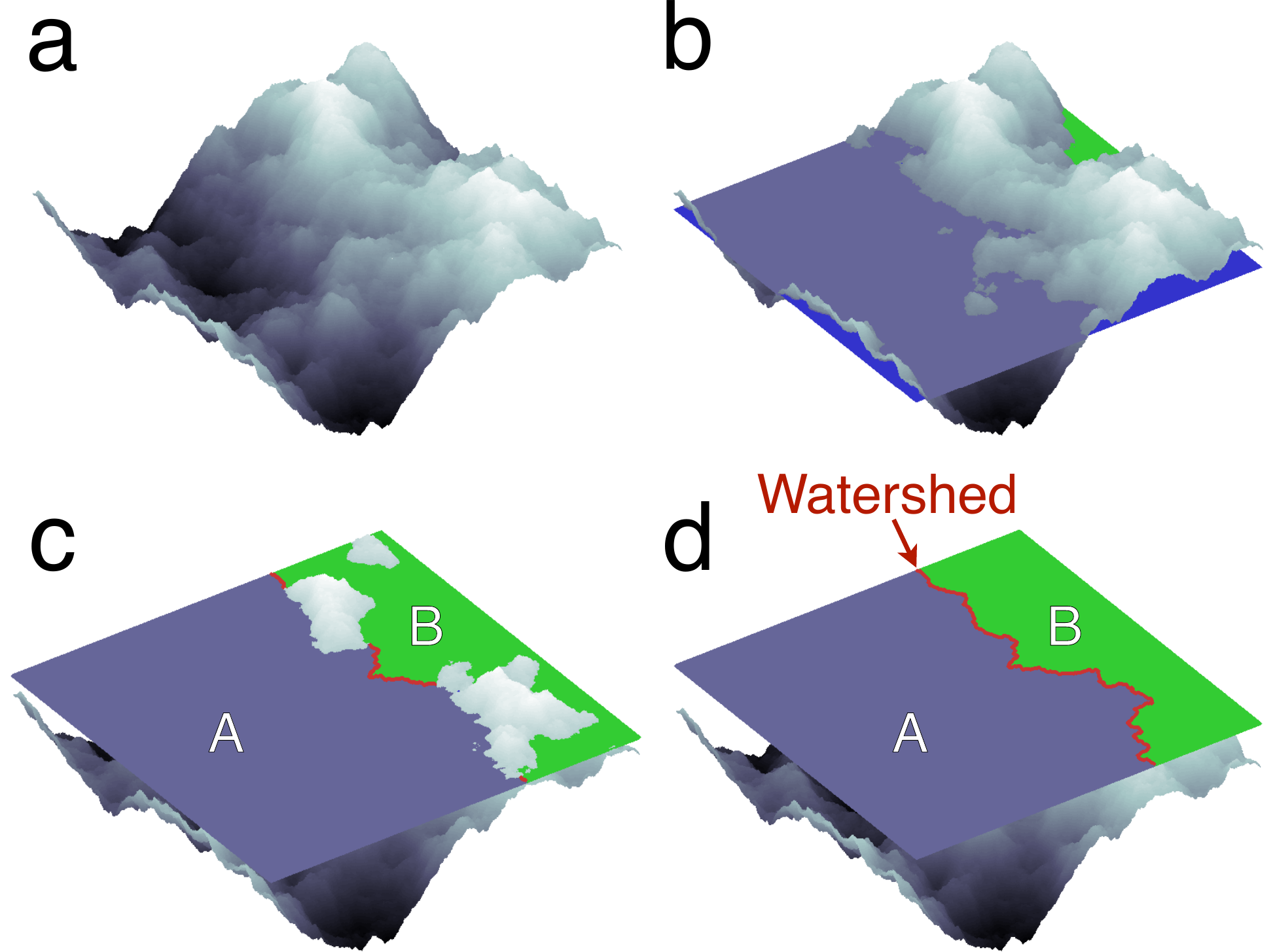}
\end{center}
\caption{Watershed dividing a landscape into two parts. The landscape
(a) is flooded from the valleys such that all regions lower than a certain
height are covered with water (b). As the water level rises, lakes merge under
the constraint that no lake can connect two predefined opposite boundaries of
the landscape (c). In this example, we have taken the lower-left and the
upper-right boundaries.  Thus, a {\it watershed} line emerges separating the
two final lakes (d).  These lakes (A and B) are the drainage basins of this
landscape.~\label{fig::watershedflood}}
\end{figure}

Since watersheds provide information about the dynamics of surface
water, they play a fundamental role in water
management~\cite{Vorosmarty98,Kwarteng00,Sarangi05}, landslides
\cite{Dhakal04,Pradhan06,Lazzari06,Lee06}, and flood prevention
\cite{Lee06,Burlando94,Yang07}. Watersheds are also of relevance in the
political context they have been used to demarcate borders between
countries such as the one between Argentina and Chile~\cite{UN02}. Thus,
the understanding of their statistical properties and resilience to
changes in the topography of the landscape are two important questions
that we will review here. In general, for every landscape, several
outlets can be defined, each one with a corresponding drainage basin.
Sets of small drainage basins eventually drain towards the same outlet
forming a even larger basin. Such hierarchy results in a larger number
of watersheds.  However, without loss of generality, the study of
watershed lines can be simplified by splitting the landscape into only
two large basins, each one draining towards opposite boundaries.
Figure~\ref{fig::watershedflood} shows how this watershed can be
identified by flooding the landscape from the valleys (see supplemental
video). Two sinks are initially defined (the lower-left and upper-right
boundaries in the example).  While flooding the landscape, each time two
lakes (A and B) draining towards opposite sinks are about to connect,
one imposes a physical barrier between the two. In the end, the
watershed line is the line formed by the barriers that separate these
two lakes.

The concept of watersheds is also of interest in other fields like,
e.g., in medical image processing~\cite{Dixon79}. There, computed
tomography scans need to be segmented to identify different tissues.
The pictures are discretized into pixels and a number is assigned to
each one of them according to the intensity. The segmentation procedure
consists in clustering neighboring pixels following the order of
increasing intensity gradient, splitting the image into different parts
(tissues).  The equivalent to the watershed line corresponds to the line
separating two different tissues~\cite{Yan06,Patil09}.

It was recently shown that watersheds can be described in the context of
percolation theory in terms of bridges and cutting bond
models~\cite{Schrenk12}, with numerical evidence that they are Schramm-Loewer
Evolution (SLE) curves in the continuum limit~\cite{Daryaei12}. This
association explains why watersheds on uncorrelated landscapes as well as
other statistical physics models, such as, optimal path
cracks~\cite{Andrade09,Andrade11,Oliveira11}, fuse networks~\cite{Moreira12},
and loopless percolation~\cite{Schrenk12}, belong to the same universality
class of optimal paths in strongly disordered media.

\section{The landscape}\label{subsec::landscape}
\begin{figure}
\begin{center}
\includegraphics[width=\columnwidth]{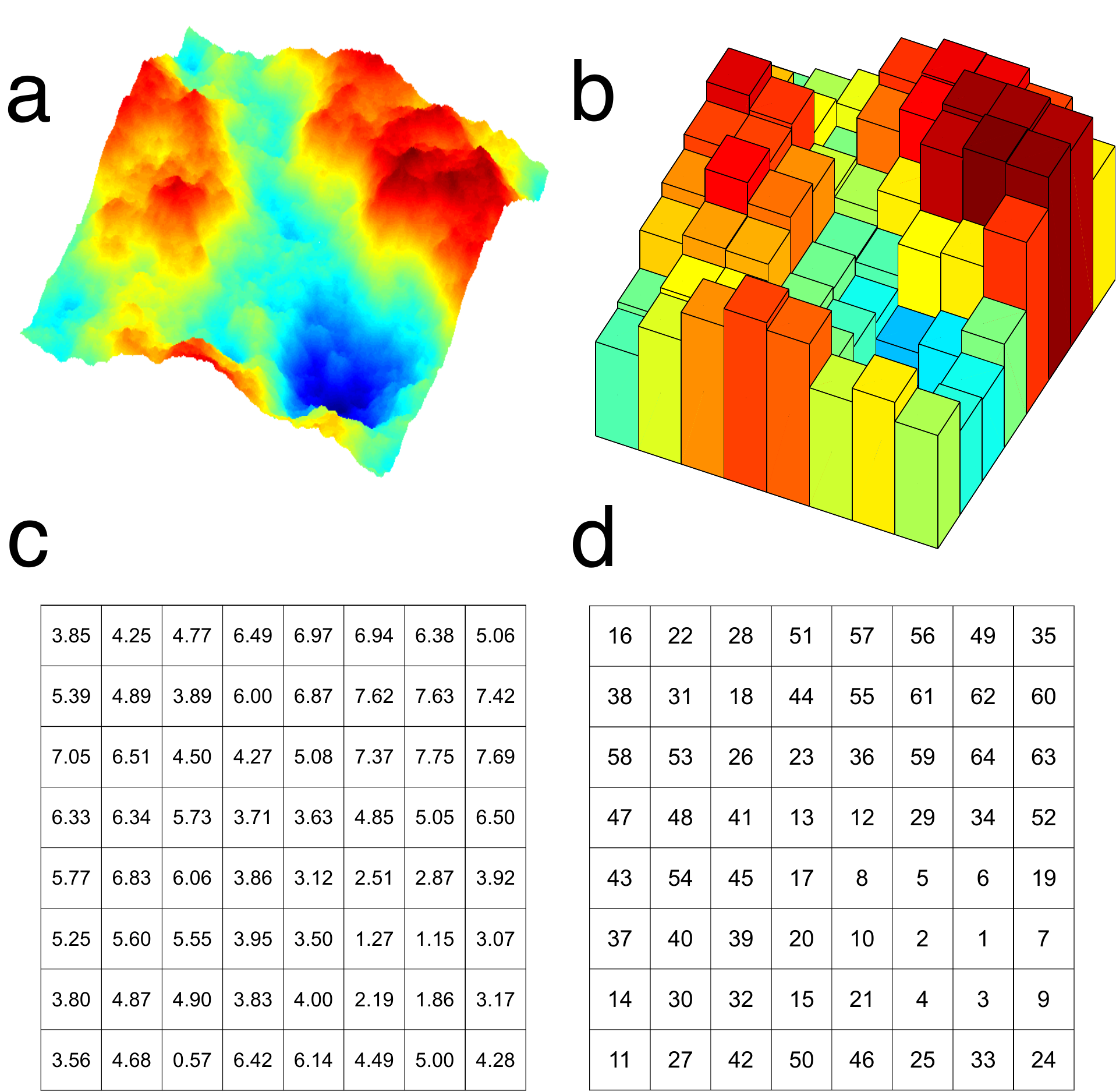}
\end{center}
\caption{The
process of generating a ranked surface. The landscape in (a) is coarse-grained
to the low-resolution system of $8 \times 8$ cells shown in (b), and then
represented as a discretized map of local heights, as depicted in (c).  By
ranking these heights in crescent order, one obtains the ranked surface in (d).
In fact, the landscape shown in (a) is a high resolution synthetic map obtained
from a fractional Brownian motion simulation based on the Fourier filtering
method (see Sec.~\ref{sec::longrange}).~\label{fig::rank}}
\end{figure}
The motion of surface water is defined by the topology of the
landscape, usually characterized by the spatial distribution of heights.
Although this distribution is a continuous function for real landscapes,
it is typically coarse-grained and represented as a digital elevation
map (DEM) of regular cells (e.g., a square lattice of sites or bonds) to
which average heights can be associated. This process is exemplarily
shown in Figs.~\ref{fig::rank}(a)-(c). In fact, modern procedures of
analyzing real landscapes numerically process
Grayscale Digital Images where the gray intensity of each pixel is
transformed into a height, resulting in a DEM. As such,
discretized maps have been useful to delimit spatial boundaries in a
wide range of problems, from tracing watersheds and river networks in
landscapes~\cite{Stark91,Maritan96,Manna95,Knecht11,Baek11} to the
identification of cancerous cells in human tissues~\cite{Yan06,Ikedo07},
and the study of spatial competition in multispecies ecosystems
\cite{Kerr06,Mathiesen11}. 

To study watersheds theoretically one can generate \textit{artificial
landscapes}. Starting with a regular lattice, a numerical value is
randomly assigned to each element, corresponding to its average height. If these
values are spatially correlated, a \textit{correlated artificial
landscape} is obtained. Otherwise, the landscape is called an
\textit{uncorrelated artificial landscape}. Natural landscapes are
characterized by long-range correlations~\cite{Pastor-Satorras98}.
In Sec.~\ref{sec::longrange} we will review how to generate correlated
artificial landscapes and their main properties.

The DEM can be further simplified by mapping them onto ranked surfaces,
where every element (site or bond) has a unique rank associated with its
corresponding value~\cite{Schrenk12}. A ranked surface can be defined in
the following way. Given a two-dimensional discretized map of size $L
\times L$, one generates a list containing the heights of its elements
(sites or bonds) in crescent order, and then replaces the numerical
values in the original map by their corresponding (integer) ranks. As
depicted in Fig.~\ref{fig::rank}(d), the result is a ranked surface. 

\section{How to determine the watershed}\label{subsec::detwatersheds}

Traditional cartographic methods for basin delineation have relied on
manual estimation from iso-elevation lines. Instead, modern procedures
are now based upon automatic processing of images as the ones obtained
from satellites~\cite{Farr07,Vincent91}.  These images are typically
coarse-grained into discretized maps with each cell $i$ having a height
$h_i$ as explained in Sec.~\ref{subsec::landscape}. A recently proposed
algorithm to identify watersheds that became rather popular consists in
the \textit{flooding procedure} described in
Fig.~\ref{fig::watershedflood}~\cite{Vincent91}, considering two sinks
such as, for example, two opposite boundaries.  Cells in the discretized
map are ranked according to their height, which leads to a ranked
surface, and are sequentially occupied, following the rank, from the
lowest to the highest. Neighboring occupied cells are then connected and
considered part of the same drainage basin, except when during this
process their connection would promote the agglomeration of the two
basins draining towards different sinks. In this case, their connection
is avoided, since they should belong to different basins.  The edge
between them is part of the watershed and, at the end, the set of such
edges forms one single watershed line that splits the landscape into two
drainage basins (see Fig.~\ref{fig::watershedflood}).
\begin{figure}
\begin{center}
\includegraphics[width=\columnwidth]{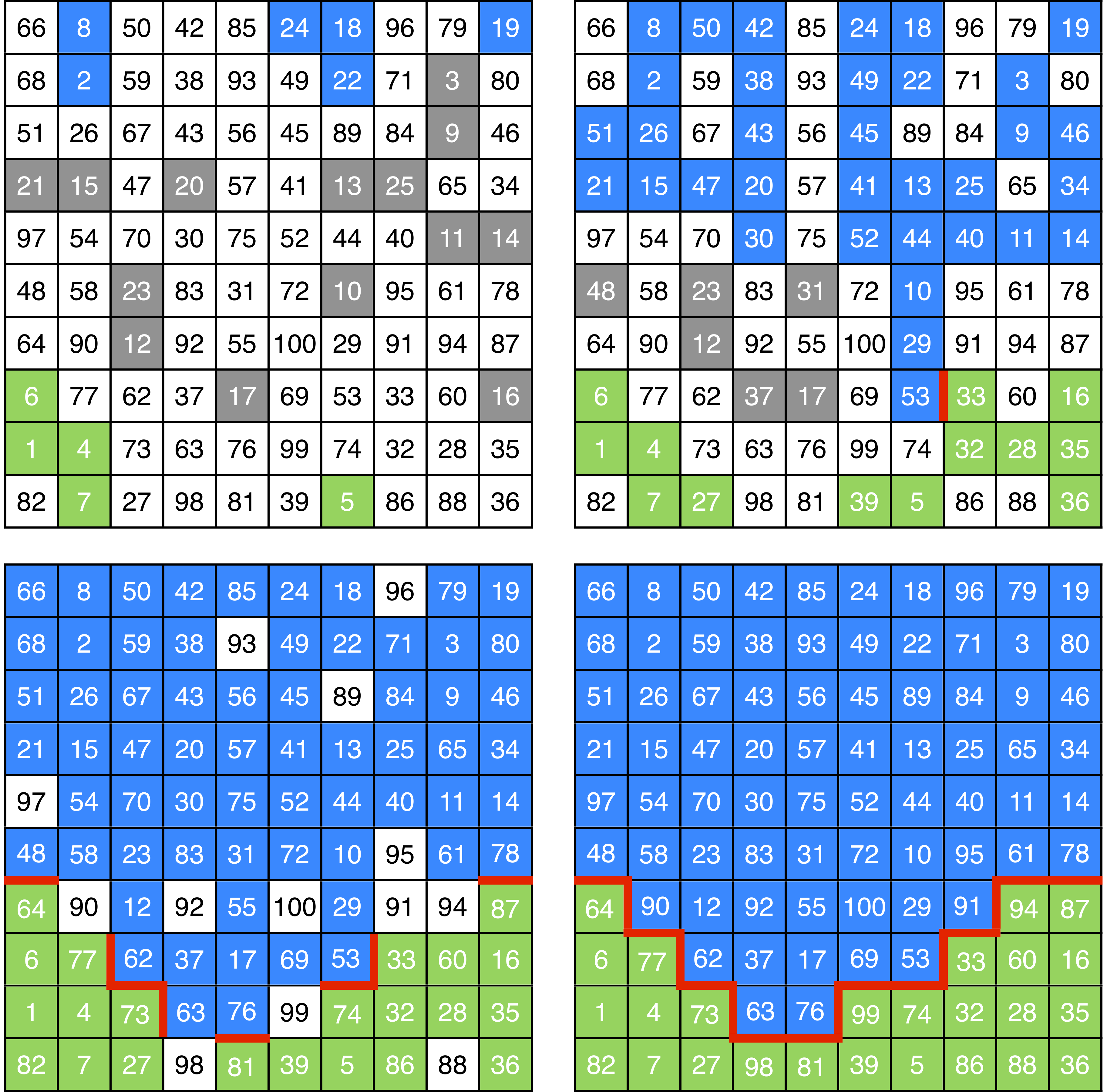}
\end{center}
\caption{Flooding algorithm to determine the watershed. Two sinks are
considered: the top row and the bottom one. Cells in the discretized map are
occupied following the rank, from the lowest to the highest.  Neighboring
occupied cells are considered connected and part of the same basin (gray
cells). The basins connected to the top row (blue) correspond to the ones
draining towards the top sink. Similarly, the ones connected to the bottom row
(green) drain towards the bottom sink (upper-left panel). If two neighbors of
the next cell in the rank belong to basins of different sinks, the edges in
contact with the cells are marked as elements of the watershed (red-thick
edges). The process proceeds iteratively until all cells are flooded (bottom
panels). At the end (bottom-right panel), the watershed splits the discretized
map into two basins (blue and green).~\label{fig::watershedalgflood}}
\end{figure}

\begin{figure}
\begin{center}
\includegraphics[width=\columnwidth]{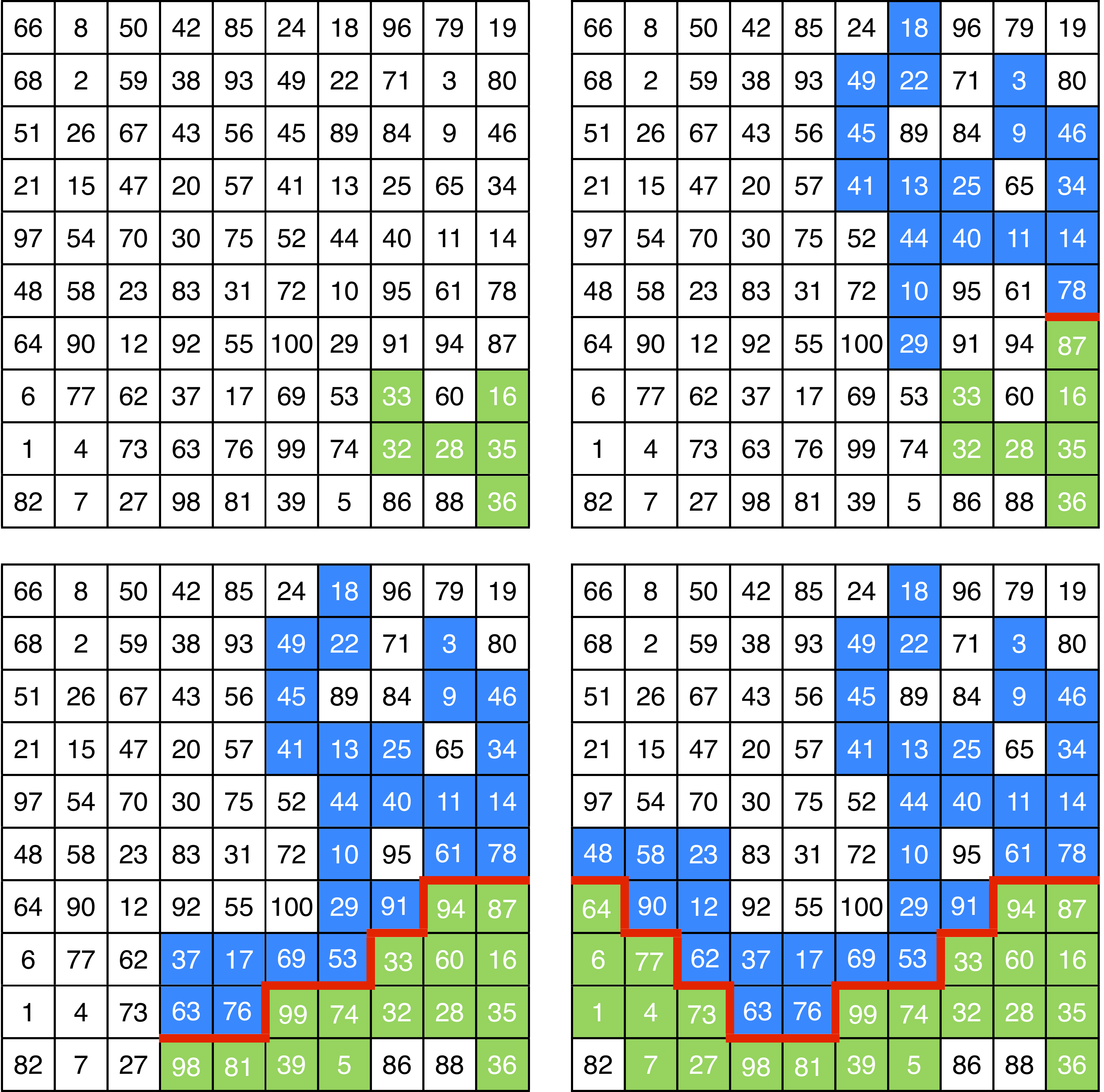}
\end{center}
\caption{Algorithm based on Invasion Percolation. As in
Fig.~\ref{fig::watershedalgflood}, two sinks are considered: the top and the
bottom rows. One starts from the bottom-right corner (cell $36$).  Since this
site is in the bottom row, it is already part of the bottom sink (green).
Proceeding upwards, following the most right column, the next unexplored cell
is considered (cell $35$). The basin to which this cell belongs grows by adding
the smallest-height cell on its perimeter, until one sink is reached (cells
$16$, $28$, $32$, and $33$ in the upper-left panel, reaching $36$).  Proceeding
further to the top, the cell $87$ is added to the (green) basin (connecting to
$16$). When the cell $78$ is considered and the corresponding basin grown, a
new (blue) basin is found draining towards the upper sink (upper-right panel).
The cell-edge between cell $87$ and $78$ is considered part of the watershed
(red-thick edge). Proceeding by iteratively identifying the basin of the cells
in the neighborhood of the watershed edges, the full watershed is identified
(bottom panels).  Since the same initial ranked surface is considered, the
final watershed is equal to the one obtained in
Fig.~\ref{fig::watershedalgflood}.~\label{fig::watershedalgip}}
\end{figure}

The flooding procedure implies visiting every cell in the discretized
map. Fehr and co-workers~\cite{Fehr09} have devised a more efficient
identification algorithm where only a fractal subset needs to be
visited. The algorithm is based on Invasion Percolation
(IP)~\cite{Wilkinson83} and consists in the following procedure. One
initially defines two sinks (for example, the bottom and top rows of the
discretized map) and considers that the (non-invaded) cell with the
lowest height on the perimeter of the (already) invaded region is the
next cell to be invaded.  If when starting from one cell a sink is
invaded, the initial cell and the entire invaded region is considered to
belong to the catchment basin of that sink. Consider that the invasion
is initially started from the cell in the bottom-right corner. Since
this cell is part of the basin of the bottom sink, the invaded region
has only one cell.  One proceeds with a new invasion from the next cell
upwards (cell $35$ in Fig.~\ref{fig::watershedalgip}) and a new invasion
cluster is grown until a sink is reached. Sequentially, all other cells
are considered in the same way. The watershed is then identified as the
line splitting the landscape into two catchment basins. The efficiency
is significantly improved if, once the first cell-edge belonging to the
watershed has been identified, the exploration continues from the cells
in the neighborhood of this edge (see Fig.~\ref{fig::watershedalgip}).
Thus, instead of visiting all cells, only a subset of $N_\mathrm{exp}$
cells needs to be explored. For uncorrelated random landscapes, the size
of this subset scales with the linear size of the DEM $L$ as
$N_\mathrm{exp}\sim L^D$, with $D=1.8\pm0.1$~\cite{Fehr09}.

\section{Fractal dimension}\label{subsec::fractaldim}
\begin{table}
\caption{\label{table::watershedTab_Df}
Fractal dimension of watersheds for natural landscapes
obtained from satellite imagery~\cite{Farr07} in Ref.~\cite{Fehr11b}. The
values for the Alps and Himalayas are from Ref.~\cite{Fehr09}. The error bars
are of the order of $2\%$.}

\begin{tabular}{ll}
\hline
\hline
Landscape & $d_f$\\
\hline
Alpes & 1.10\\
Europe & 1.10\\
Rocky Mountains & 1.11\\
Himalayas & 1.11\\
Kongo & 1.11\\
Andes & 1.12\\
Appalachians & 1.12\\
Brazil & 1.12\\
Germany & 1.14\\
Big Lakes & 1.15\\
\hline
\hline
\end{tabular}
\end{table}

Breyer and Snow~\cite{Breyer92} have studied $12$ basins in the United
States and concluded that their watershed lines are self-similar
objects~\cite{Mandelbrot67,Mandelbrot83}. A self-similar structure is
characterized by its fractal dimension $d_f$, which is defined here
through the scaling of the number $M$ of edges belonging to the
watershed with the linear size $L$ of the discretized map,
\begin{equation}\label{EQfractaldim}
M\sim L^{d_f}\,\, .
\end{equation}
They obtained fractal dimensions in the range $1.05-1.12$.  Fehr and
co-workers~\cite{Fehr09}, with the method described in
Sec.~\ref{subsec::detwatersheds} confirmed this self-similar behavior
over more than three orders of magnitude and measured the fractal
dimension for watersheds in several real landscapes, as summarized in
Tab.~\ref{table::watershedTab_Df}. 

For uncorrelated artificial landscapes the watershed fractal dimension
has been estimated to be
\mbox{$d_f=1.2168\pm0.0005$}~\cite{Cieplak94,Fehr09,Fehr11c}. This value
has drawn considerable attention since it was also found in several
other physical models such as optimal paths in strong disorder and optimal path
cracks~\cite{Porto97,Buldyrev04,Andrade09,Andrade11,Oliveira11}, bridge
percolation \cite{Cieplak94,Cieplak96,Schrenk12}, and the surface of
explosive percolation clusters \cite{Araujo10,Schrenk11}. The conjecture
that all these models might belong to the same universality class has
opened a broad range of possible implications and applications of the
properties of watersheds. As discussed in Ref.~\cite{Schrenk12}, the
relation between most of such models can be established when they are
described within the framework of ranked surfaces (see also
Sec.~\ref{subsec::landscape}).

\section{Watersheds in three and higher dimensions}

Up to now, we have focused on the watershed line that divides the
landscape into drainage basins for water on the surface. However, in
reality, water also penetrates the soil and flows underground. Thus, the
concept of watersheds and ranked surfaces can be extended to three
dimensions. The soil can be described as a porous medium consisting of a
network of pores connected through channels. When a fluid penetrates
through this medium, a threshold pressure $p_k$ can be defined for each
channel $k$ such that the channel can only be invaded when $p\geq p_k$,
where $p$ is the fluid pressure. In general, a channel $k$ is closed
when $p_k>p$, and open otherwise. This system can be mapped into a three
dimensional ranked volume, where the lattice elements are the channels
and the rank is defined by the increasing order of the threshold pressure. The
sequence in the rank corresponds to the order of channel openings when
the fluid pressure is quasistatically raised from zero. Analogously to the
ranked surfaces, one can split the space in two regions,
draining towards opposite boundaries.

\begin{figure}
\begin{center}
\includegraphics[width=\columnwidth]{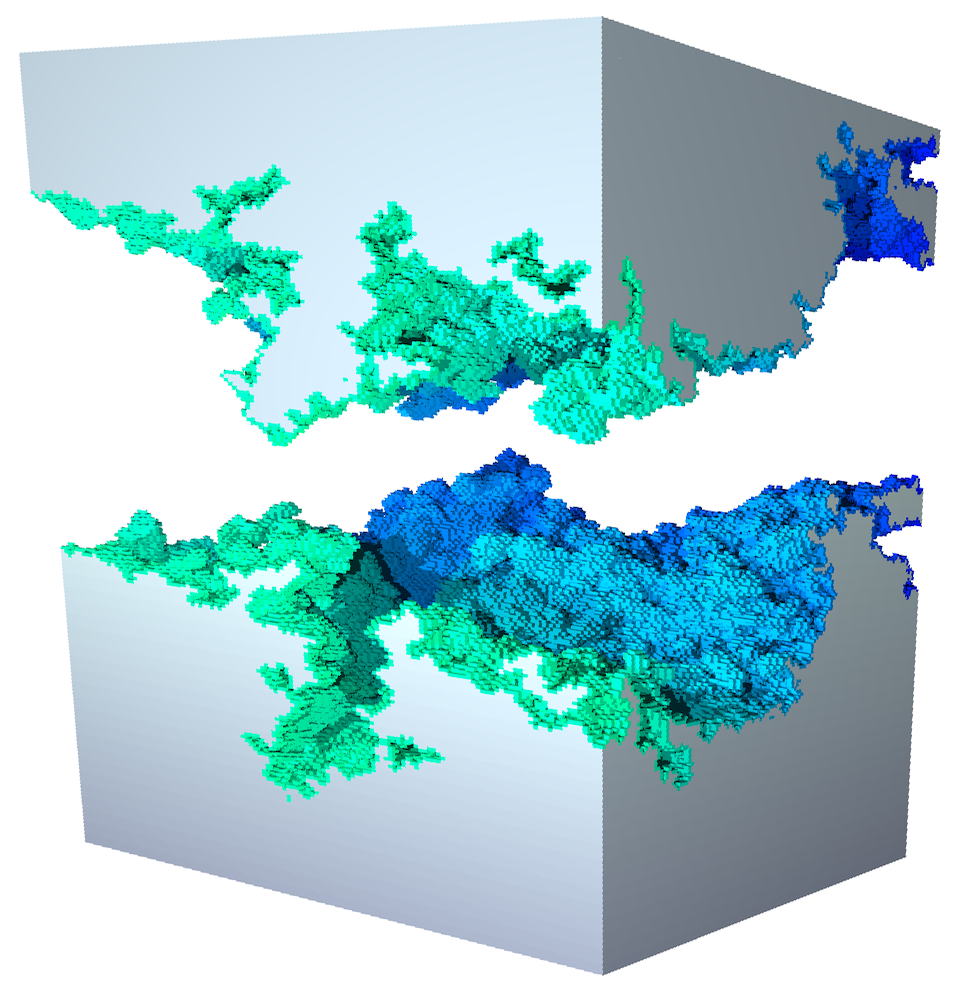}
\end{center}
\caption{Watershed in three dimensions. Example of a watershed for an
uncorrelated three-dimensional space, namely, a simple-cubic lattice of $128^3$
sites. To obtain the watershed, sites are sequentially occupied under the
constraint that groups of connected sites in contact with the top boundary
cannot merge with the ones connected to the bottom boundary. In three
dimensions, the fractal dimension is
$d_f=2.487\pm0.003$~\cite{Fehr11c}.~\label{fig::watershed3d}}
\end{figure}

The watershed in three dimensions is a surface of fractal dimension
$d_f=2.487\pm0.003$~\cite{Fehr11c}. An example for a simple-cubic ranked
volume is shown in Fig.~\ref{fig::watershed3d}. In general, for lattices
of size $L^d$, where $d$ is the spatial dimension, the watershed
blocks connectivity from one side to the other. Thus, the watershed
fractal dimension must follow $d-1\leq d_f\leq d$, i.e., with increasing
dimension $d_f$ also increases~\cite{Schrenk12}. 

\section{Impact of perturbation on
watersheds}\label{subsec::impact_perturb}

The stability of watersheds is also a subject of interest. For example,
changes in the watershed might affect the sediment supply of
rivers~\cite{Douglass07}. Also, the understanding of the temporal
evolution of drainage networks provides valuable insight into the
biodiversity between basins~\cite{Bishop95}. Geographers and
geomorphologists have found that the evolution of watersheds is typically
driven by local changes of the landscape~\cite{Burridge07}. These
events can be triggered by various mechanisms like
erosion~\cite{Garcia-Castellanos09,Linkeviciene09,Bishop95}, natural
damming~\cite{Lee06}, tectonic motion
\cite{Garcia-Castellanos03,Dorsey06,Lock06}, as well as volcanic
activity~\cite{Beranek06}. 
Although rare, these local events can have
a huge impact on the hydrological
system~\cite{Attal08,Garcia-Castellanos09,Linkeviciene09,Burridge07}.
For example, it was shown that a local height
change of less than two meters at a location close to the Kashabowie
Provincial Park, some kilometers North of the US-Canadian
border, can trigger a displacement in the watershed such that the area
enclosed by the original and the new watersheds is about
$3730~\mbox{km}^2$~\cite{Fehr11}. 

The stability of watersheds is also relevant in the political context.
For example, Chile and Argentina share a common border with more than
$5000$ kilometers, which was the source of a long dispute between these
countries~\cite{UN02}. A treaty established this border as being the
watershed between the Atlantic and Pacific Oceans for several segments.
In $1902$, the Argentinian Francisco Moreno contributed significantly to
elucidate the technical basis for dispute. He proved that during the
quarternary glaciations, the watershed line changed. In particular,
several Patagonian lakes currently draining to the Pacific Ocean were in
fact originally part of the Atlantic Ocean basin.  Consequently, he
argued, instead of belonging to Chile they should be awarded to
Argentina.

Douglas and Schmeeckle~\cite{Douglass07} have performed fifteen table
top experiments to study the mechanisms of drainage rearrangements.  In
spite of being diverse, the mechanisms triggering the evolution of
watersheds are all modifications of the
topography~\cite{Bishop95,Lee06,Garcia-Castellanos03}. In that
perspective, the effect of such events can be investigated by applying
small local perturbations to natural and artificial landscapes and
analyzing the changes in the watershed~\cite{Fehr11,Fehr11b}.
Specifically, one starts with a discretized landscape and computes the
original watershed. A local event is then induced by changing the height
of a site $k$ $h_k\rightarrow h_k+\Delta$, where $\Delta$ is the
perturbation strength, and the new watershed is identified. The impact
of perturbations on watersheds in two and three dimensions will be
discussed in the following.

\subsection{Impact of perturbations in two dimensions}\label{subsec::pert2d}

Fehr and co-workers \cite{Fehr11} have normalized the perturbation strength
by the height difference between the highest and lowest height of the
landscape. For each landscape, they have sequentially perturbed every
site of the corresponding discretized map, such that each perturbed
landscape differs from the original only in one single site. The
effect of those perturbations affecting the watershed has been
quantified by the properties of the region enclosed by the original and
the new watersheds.  For that region, they have measured its area,
corresponding to the number of sites, $N_s$, in the discretized map and
the distance $R$ between its original and new {\it outlets}, defined as
the points where water escapes from this region.  Note that, there are
only two outlets involved in this procedure: one related to the original
landscape and the other to the new one. Since $\Delta$ is strictly
positive, the outlet in the new landscape corresponds always to the
perturbed site $k$.

Scale-free behavior has been found for the distribution $P(N_s)$ of the
number of enclosed sites $N_s$, the probability distribution $P(R)$ of
the distance $R$ between outlets, and the dependence of the average
$\langle N_s\rangle$ on $R$~\cite{Fehr11,Fehr11b}. Specifically,
\begin{subequations}
\begin{eqnarray}
P(N_s) &\sim& N_s^{-\beta},\label{EQ_PA}\\
P(R) &\sim& R^{-\rho},\label{EQ_PR}\\
\langle N_s\rangle &\sim& R^{\sigma},\label{EQ_AR}
\end{eqnarray}
\end{subequations}
where the measured exponents are summarized in
Tab.~\ref{tab::impact_exp}. The power-law decay (\ref{EQ_PR}) and the
relation (\ref{EQ_AR}) imply that a localized perturbation can have a
large impact on the shape of the watershed even at very large distances,
hence having a non-local effect. Additionally, the analysis
of the fraction of perturbed sites affecting the watershed revealed a
power-law scaling with the strength $\Delta$. This finding supports the
conclusion that changes in the watershed can be even triggered by
anthropological small perturbations~\cite{Fehr11}.

\begin{table}[!t]
\caption{\label{tab::impact_exp}
Exponents for the impact of perturbations on
watersheds, calculated for uncorrelated artificial landscapes in two
and three dimensions, from
Refs.~\cite{Stauffer94,Fehr11,Fehr11b,Fehr11c}.}

\begin{tabular}{cr@{$\pm$}lr@{$\pm$}l}\toprule
Exponent &\multicolumn{4}{c}{Uncorrelated landscapes}\\
\hline
$d$&\multicolumn{2}{c}{2}&\multicolumn{2}{c}{3}\\
$d_f$&1.2168&0.0005&2.487&0.003\\
$\beta$&1.16&0.03&1.31&0.05\\
$\rho$&2.21&0.01&3.2&0.2\\
$\sigma$&\multicolumn{2}{c}{2}&2.45&0.05\\
$\alpha^*$&1.39&0.03&1.4&0.1\\ \botrule
\end{tabular}
\end{table}

For the region enclosed by the original and the new watersheds, an invasion
percolation (IP) cluster can be obtained by imposing a pressure drop between the
outlet in the new watershed and the one in the original one, always invading
along the steepest descent of the entire cluster perimeter. The size
distribution $P(M_{\mathrm{IP}}|R)$ of these clusters, for each fixed distance
$R$ between outlets, has been shown to scale as,
\begin{equation}\label{eq::pertipclust} P(M_{\mathrm{IP}}|R)\sim
M_{IP}^{-(1+\alpha^*)} \ \ , \end{equation}
where $\alpha^*\approx 1.39$. This exponent corresponds to the one found for
the size distribution of point-to-point IP-clusters~\cite{Araujo05}.  In that
process, invasion clusters are obtained in a random medium by invading from one
point to another at a certain distance. By contrast, in the watershed case the
invasion is always started from the outlet on the new watershed. This
difference between starting at any point or in the outlet justifies the
additional factor of unity in the scaling (Eq.~(\ref{eq::pertipclust})), since
the probabilities need to be rescaled by the size of the
IP-cluster~\cite{Fehr11}.

\subsection{Impact of perturbations in three dimensions}\label{subsec::pert3d}

The impact of perturbations on watersheds has been also analyzed in three
dimensions \cite{Fehr11b}. Similar to 2D, a perturbation is induced by changing
the value of a site $k$, $h_k\rightarrow h_k+\Delta$, where $\Delta$ is the
perturbation strength. The number of sites $N_s$ enclosed by the original and
new watersheds corresponds to a volume. Power-law scaling in terms of
Eqs.~(\ref{EQ_PA})-(\ref{EQ_AR}) has also been observed with the exponents
summarized in Tab.~\ref{tab::impact_exp}. The value of $\alpha^*$ is similar to
the one found in two dimensions. According to Lee~\cite{Lee09}, the size
distribution of the point-to-point IP-cluster is independent on the
dimensionality of the system. Therefore, the numerical agreement between
$\alpha^*$ for different spatial dimensions supports the relation with invasion
percolation. 

\section{Watersheds on long-range correlated landscapes}\label{sec::longrange}

Results discussed heretofore were obtained on random uncorrelated
landscapes. However real landscapes are characterized by spatial long-range correlated
height distributions. Numerically, such distributions can be generated from
fractional Brownian motion (fBm) \cite{Mandelbrot67,Peitgen88}, using the
Fourier filtering method
\cite{Fehr11,Morais11,Sahimi94a,Sahimi96,Makse96,Oliveira11,
Prakash92,Kikkinides99,Stanley99,Makse00,Araujo02,Araujo03,Du04}. This method
allows to control the nature and the strength of correlations, which are
characterized by the Hurst exponent $H$. The uncorrelated distribution of
heights is solely obtained for $H=-d/2$, i.e., $H=-1$ and $H=-3/2$ in two and
three dimensions, respectively. A detailed description of this method can be
found, e.g., in Ref.~\cite{Oliveira11,Peitgen88}. 

Fehr et al.~\cite{Fehr11,Fehr11b} used fractional Brownian motion (fBm) on a
square lattice \cite{Lauritsen93} to incorporate long-range correlations
controlled by the Hurst exponent $H$.  They calculated how the fractal
dimension decreases with the Hurst exponent and found good quantitative
agreement with the exponents obtained for natural landscapes, typically with
$0.3<H<0.5$, which is the known range of Hurst exponents for real landscapes on
length scales larger than 1 km (see Ref.~\cite{Pastor-Satorras98} and
references therein). They also obtained $\alpha$, $\beta$ and $\rho$ for
several values of $H$, observing that both $\beta$ and $\rho$ increase with
$H$, finding also good quantitative agreement. Thus, their model provides a
complete quantitative description of the effects observed on natural
landscapes.

\section{Final Remarks}
Here we solely discussed cases with one watershed, but the same
theoretical framework can be straightforwardly extended to tackle other
space partition problems. For example, the identification of the entire
set of watersheds of a landscape with multiple outlets helps identifying
the catchment areas contributing to each river or
reservoir~\cite{Mamede12}.  A systematic study of disordered media with
multiple outlets is still missing. Examples of open questions are: How
does the distribution of catchment basins or the number of triplets
(points where two watersheds meet) depend on the number of outlets? And,
how does the statistics of perturbations change in the presence of
triplets? The division of a volume into several parts is also a problem
of practical interest in the extraction of resources from porous
soils~\cite{Schrenk12b}. Studies of such systems have mainly considered
uncorrelated disordered media. The role of long-range correlation there
is still an open problem.

\section*{Acknowledgement}

We acknowledge financial support from the
European Research Council (ERC) Advanced Grant 319968-FlowCCS, the Brazilian
Agencies CNPq, CAPES, FUNCAP and FINEP, the FUNCAP/CNPq Pronex grant, the
National Institute of Science and Technology for Complex Systems in Brazil, the
Portuguese Foundation for Science and Technology (FCT) under contracts no.
IF/00255/2013, PEst-OE/FIS/UI0618/2014, and EXCL/FIS-NAN/0083/2012, and the
Swiss National Science Foundation under Grant No. P2EZP2-152188.

\bibliography{review}

\begin{thebibliography}{73}%
\makeatletter
\providecommand \@ifxundefined [1]{%
 \@ifx{#1\undefined}
}%
\providecommand \@ifnum [1]{%
 \ifnum #1\expandafter \@firstoftwo
 \else \expandafter \@secondoftwo
 \fi
}%
\providecommand \@ifx [1]{%
 \ifx #1\expandafter \@firstoftwo
 \else \expandafter \@secondoftwo
 \fi
}%
\providecommand \natexlab [1]{#1}%
\providecommand \enquote  [1]{``#1''}%
\providecommand \bibnamefont  [1]{#1}%
\providecommand \bibfnamefont [1]{#1}%
\providecommand \citenamefont [1]{#1}%
\providecommand \href@noop [0]{\@secondoftwo}%
\providecommand \href [0]{\begingroup \@sanitize@url \@href}%
\providecommand \@href[1]{\@@startlink{#1}\@@href}%
\providecommand \@@href[1]{\endgroup#1\@@endlink}%
\providecommand \@sanitize@url [0]{\catcode `\\12\catcode `\$12\catcode
  `\&12\catcode `\#12\catcode `\^12\catcode `\_12\catcode `\%12\relax}%
\providecommand \@@startlink[1]{}%
\providecommand \@@endlink[0]{}%
\providecommand \url  [0]{\begingroup\@sanitize@url \@url }%
\providecommand \@url [1]{\endgroup\@href {#1}{\urlprefix }}%
\providecommand \urlprefix  [0]{URL }%
\providecommand \Eprint [0]{\href }%
\providecommand \doibase [0]{http://dx.doi.org/}%
\providecommand \selectlanguage [0]{\@gobble}%
\providecommand \bibinfo  [0]{\@secondoftwo}%
\providecommand \bibfield  [0]{\@secondoftwo}%
\providecommand \translation [1]{[#1]}%
\providecommand \BibitemOpen [0]{}%
\providecommand \bibitemStop [0]{}%
\providecommand \bibitemNoStop [0]{.\EOS\space}%
\providecommand \EOS [0]{\spacefactor3000\relax}%
\providecommand \BibitemShut  [1]{\csname bibitem#1\endcsname}%
\let\auto@bib@innerbib\@empty
\bibitem [{\citenamefont {Gregory}\ and\ \citenamefont
  {Walling}(1973)}]{Gregory73}%
  \BibitemOpen
  \bibfield  {author} {\bibinfo {author} {\bibfnamefont {K.~J.}\ \bibnamefont
  {Gregory}}\ and\ \bibinfo {author} {\bibfnamefont {D.~E.}\ \bibnamefont
  {Walling}},\ }\href@noop {} {\emph {\bibinfo {title} {Drainage basin form and
  process: {A} geomorphological approach}}}\ (\bibinfo  {publisher} {Edward
  Arnold},\ \bibinfo {address} {London},\ \bibinfo {year} {1973})\BibitemShut
  {NoStop}%
\bibitem [{\citenamefont {V\"or\"osmarty}\ \emph {et~al.}(1998)\citenamefont
  {V\"or\"osmarty}, \citenamefont {Federer},\ and\ \citenamefont
  {Schloss}}]{Vorosmarty98}%
  \BibitemOpen
  \bibfield  {author} {\bibinfo {author} {\bibfnamefont {C.~J.}\ \bibnamefont
  {V\"or\"osmarty}}, \bibinfo {author} {\bibfnamefont {C.~A.}\ \bibnamefont
  {Federer}}, \ and\ \bibinfo {author} {\bibfnamefont {A.~L.}\ \bibnamefont
  {Schloss}},\ }\href {\doibase 10.1016/S0022-1694(98)00109-7} {\bibfield
  {journal} {\bibinfo  {journal} {J. Hydrol.}\ }\textbf {\bibinfo {volume}
  {207}},\ \bibinfo {pages} {147} (\bibinfo {year} {1998})}\BibitemShut
  {NoStop}%
\bibitem [{\citenamefont {Kwarteng}\ \emph {et~al.}(2000)\citenamefont
  {Kwarteng}, \citenamefont {Viswanathan}, \citenamefont {Al-Senafy},\ and\
  \citenamefont {Rashid}}]{Kwarteng00}%
  \BibitemOpen
  \bibfield  {author} {\bibinfo {author} {\bibfnamefont {A.~Y.}\ \bibnamefont
  {Kwarteng}}, \bibinfo {author} {\bibfnamefont {M.~N.}\ \bibnamefont
  {Viswanathan}}, \bibinfo {author} {\bibfnamefont {M.~N.}\ \bibnamefont
  {Al-Senafy}}, \ and\ \bibinfo {author} {\bibfnamefont {T.}~\bibnamefont
  {Rashid}},\ }\href@noop {} {\bibfield  {journal} {\bibinfo  {journal} {J.
  Arid. Environ.}\ }\textbf {\bibinfo {volume} {46}},\ \bibinfo {pages} {137}
  (\bibinfo {year} {2000})}\BibitemShut {NoStop}%
\bibitem [{\citenamefont {Sarangi}\ and\ \citenamefont
  {Bhattacharya}(2005)}]{Sarangi05}%
  \BibitemOpen
  \bibfield  {author} {\bibinfo {author} {\bibfnamefont {A.}~\bibnamefont
  {Sarangi}}\ and\ \bibinfo {author} {\bibfnamefont {A.~K.}\ \bibnamefont
  {Bhattacharya}},\ }\href@noop {} {\bibfield  {journal} {\bibinfo  {journal}
  {Agr. Water Manage.}\ }\textbf {\bibinfo {volume} {78}},\ \bibinfo {pages}
  {195} (\bibinfo {year} {2005})}\BibitemShut {NoStop}%
\bibitem [{\citenamefont {Dhakal}\ and\ \citenamefont
  {Sidle}(2004)}]{Dhakal04}%
  \BibitemOpen
  \bibfield  {author} {\bibinfo {author} {\bibfnamefont {A.~S.}\ \bibnamefont
  {Dhakal}}\ and\ \bibinfo {author} {\bibfnamefont {R.~C.}\ \bibnamefont
  {Sidle}},\ }\href@noop {} {\bibfield  {journal} {\bibinfo  {journal} {Hydrol.
  Process.}\ }\textbf {\bibinfo {volume} {18}},\ \bibinfo {pages} {757}
  (\bibinfo {year} {2004})}\BibitemShut {NoStop}%
\bibitem [{\citenamefont {Pradhan}\ \emph {et~al.}(2006)\citenamefont
  {Pradhan}, \citenamefont {Singh},\ and\ \citenamefont
  {Buchroithner}}]{Pradhan06}%
  \BibitemOpen
  \bibfield  {author} {\bibinfo {author} {\bibfnamefont {B.}~\bibnamefont
  {Pradhan}}, \bibinfo {author} {\bibfnamefont {R.~P.}\ \bibnamefont {Singh}},
  \ and\ \bibinfo {author} {\bibfnamefont {M.~F.}\ \bibnamefont
  {Buchroithner}},\ }\href@noop {} {\bibfield  {journal} {\bibinfo  {journal}
  {Adv. Space Res.}\ }\textbf {\bibinfo {volume} {37}},\ \bibinfo {pages} {698}
  (\bibinfo {year} {2006})}\BibitemShut {NoStop}%
\bibitem [{\citenamefont {Lazzari}\ \emph {et~al.}(2006)\citenamefont
  {Lazzari}, \citenamefont {Geraldi}, \citenamefont {Lapenna},\ and\
  \citenamefont {Loperte}}]{Lazzari06}%
  \BibitemOpen
  \bibfield  {author} {\bibinfo {author} {\bibfnamefont {M.}~\bibnamefont
  {Lazzari}}, \bibinfo {author} {\bibfnamefont {E.}~\bibnamefont {Geraldi}},
  \bibinfo {author} {\bibfnamefont {V.}~\bibnamefont {Lapenna}}, \ and\
  \bibinfo {author} {\bibfnamefont {A.}~\bibnamefont {Loperte}},\ }\href@noop
  {} {\bibfield  {journal} {\bibinfo  {journal} {Landslides}\ }\textbf
  {\bibinfo {volume} {3}},\ \bibinfo {pages} {275} (\bibinfo {year}
  {2006})}\BibitemShut {NoStop}%
\bibitem [{\citenamefont {Lee}\ and\ \citenamefont {Lin}(2006)}]{Lee06}%
  \BibitemOpen
  \bibfield  {author} {\bibinfo {author} {\bibfnamefont {K.~T.}\ \bibnamefont
  {Lee}}\ and\ \bibinfo {author} {\bibfnamefont {Y.-T.}\ \bibnamefont {Lin}},\
  }\href@noop {} {\bibfield  {journal} {\bibinfo  {journal} {J. Am. Water
  Resour. As.}\ }\textbf {\bibinfo {volume} {42}},\ \bibinfo {pages} {1615}
  (\bibinfo {year} {2006})}\BibitemShut {NoStop}%
\bibitem [{\citenamefont {Burlando}\ \emph {et~al.}(1994)\citenamefont
  {Burlando}, \citenamefont {Mancini},\ and\ \citenamefont
  {Rosso}}]{Burlando94}%
  \BibitemOpen
  \bibfield  {author} {\bibinfo {author} {\bibfnamefont {P.}~\bibnamefont
  {Burlando}}, \bibinfo {author} {\bibfnamefont {M.}~\bibnamefont {Mancini}}, \
  and\ \bibinfo {author} {\bibfnamefont {R.}~\bibnamefont {Rosso}},\
  }\href@noop {} {\bibfield  {journal} {\bibinfo  {journal} {IFIP Trans. B}\
  }\textbf {\bibinfo {volume} {16}},\ \bibinfo {pages} {91} (\bibinfo {year}
  {1994})}\BibitemShut {NoStop}%
\bibitem [{\citenamefont {Yang}\ \emph {et~al.}(2007)\citenamefont {Yang},
  \citenamefont {Zhao}, \citenamefont {Armstrong}, \citenamefont {Robinson},\
  and\ \citenamefont {Brodzik}}]{Yang07}%
  \BibitemOpen
  \bibfield  {author} {\bibinfo {author} {\bibfnamefont {D.}~\bibnamefont
  {Yang}}, \bibinfo {author} {\bibfnamefont {Y.}~\bibnamefont {Zhao}}, \bibinfo
  {author} {\bibfnamefont {R.}~\bibnamefont {Armstrong}}, \bibinfo {author}
  {\bibfnamefont {D.}~\bibnamefont {Robinson}}, \ and\ \bibinfo {author}
  {\bibfnamefont {M.-J.}\ \bibnamefont {Brodzik}},\ }\href@noop {} {\bibfield
  {journal} {\bibinfo  {journal} {J. Geophys. Res.}\ }\textbf {\bibinfo
  {volume} {112}},\ \bibinfo {pages} {F02S22} (\bibinfo {year}
  {2007})}\BibitemShut {NoStop}%
\bibitem [{\citenamefont {{United Nations}}(2006)}]{UN02}%
  \BibitemOpen
  \bibfield  {author} {\bibinfo {author} {\bibnamefont {{United Nations}}},\
  }\href {http://untreaty.un.org/cod/riaa/cases/vol_IX/29-49.pdf} {\enquote
  {\bibinfo {title} {The {C}ordillera of the {A}ndes boundary case
  ({A}rgentina, {C}hile), 20 {N}ovember 1902},}\ } (\bibinfo {year} {2006}),\
  \bibinfo {note}
  {{http}://untreaty.un.org/cod/riaa/cases/vol$\_$IX/29-49.pdf}\BibitemShut
  {NoStop}%
\bibitem [{\citenamefont {Dixon}(1979)}]{Dixon79}%
  \BibitemOpen
  \bibfield  {author} {\bibinfo {author} {\bibfnamefont {J.~K.}\ \bibnamefont
  {Dixon}},\ }\href {\doibase 10.1109/TSMC.1979.4310090} {\bibfield  {journal}
  {\bibinfo  {journal} {IEEE T. Syst. Man Cyb.}\ }\textbf {\bibinfo {volume}
  {9}},\ \bibinfo {pages} {617} (\bibinfo {year} {1979})}\BibitemShut {NoStop}%
\bibitem [{\citenamefont {Yan}\ \emph {et~al.}(2006)\citenamefont {Yan},
  \citenamefont {Zhao}, \citenamefont {Wang}, \citenamefont {Zelenetz},\ and\
  \citenamefont {Schwartz}}]{Yan06}%
  \BibitemOpen
  \bibfield  {author} {\bibinfo {author} {\bibfnamefont {J.}~\bibnamefont
  {Yan}}, \bibinfo {author} {\bibfnamefont {B.}~\bibnamefont {Zhao}}, \bibinfo
  {author} {\bibfnamefont {L.}~\bibnamefont {Wang}}, \bibinfo {author}
  {\bibfnamefont {A.}~\bibnamefont {Zelenetz}}, \ and\ \bibinfo {author}
  {\bibfnamefont {L.~H.}\ \bibnamefont {Schwartz}},\ }\href {\doibase
  10.1118/1.2207133} {\bibfield  {journal} {\bibinfo  {journal} {Med. Phys.}\
  }\textbf {\bibinfo {volume} {33}},\ \bibinfo {pages} {2452} (\bibinfo {year}
  {2006})}\BibitemShut {NoStop}%
\bibitem [{\citenamefont {Patil}\ and\ \citenamefont
  {Patilkulkarani}(2009)}]{Patil09}%
  \BibitemOpen
  \bibfield  {author} {\bibinfo {author} {\bibfnamefont {C.~M.}\ \bibnamefont
  {Patil}}\ and\ \bibinfo {author} {\bibfnamefont {S.}~\bibnamefont
  {Patilkulkarani}},\ }in\ \href {\doibase 10.1109/ARTCom.2009.14} {\emph
  {\bibinfo {booktitle} {IEEE Proceedings of ARTCom 2009}}}\ (\bibinfo {year}
  {2009})\ p.\ \bibinfo {pages} {796}\BibitemShut {NoStop}%
\bibitem [{\citenamefont {Schrenk}\ \emph
  {et~al.}(2012{\natexlab{a}})\citenamefont {Schrenk}, \citenamefont
  {Ara\'ujo}, \citenamefont {{Andrade Jr.}},\ and\ \citenamefont
  {Herrmann}}]{Schrenk12}%
  \BibitemOpen
  \bibfield  {author} {\bibinfo {author} {\bibfnamefont {K.~J.}\ \bibnamefont
  {Schrenk}}, \bibinfo {author} {\bibfnamefont {N.~A.~M.}\ \bibnamefont
  {Ara\'ujo}}, \bibinfo {author} {\bibfnamefont {J.~S.}\ \bibnamefont {{Andrade
  Jr.}}}, \ and\ \bibinfo {author} {\bibfnamefont {H.~J.}\ \bibnamefont
  {Herrmann}},\ }\href {\doibase 10.1038/srep00348} {\bibfield  {journal}
  {\bibinfo  {journal} {Sci. Rep.}\ }\textbf {\bibinfo {volume} {2}},\ \bibinfo
  {pages} {348} (\bibinfo {year} {2012}{\natexlab{a}})}\BibitemShut {NoStop}%
\bibitem [{\citenamefont {Daryaei}\ \emph {et~al.}(2012)\citenamefont
  {Daryaei}, \citenamefont {Ara\'ujo}, \citenamefont {Schrenk}, \citenamefont
  {Rouhani},\ and\ \citenamefont {Herrmann}}]{Daryaei12}%
  \BibitemOpen
  \bibfield  {author} {\bibinfo {author} {\bibfnamefont {E.}~\bibnamefont
  {Daryaei}}, \bibinfo {author} {\bibfnamefont {N.~A.~M.}\ \bibnamefont
  {Ara\'ujo}}, \bibinfo {author} {\bibfnamefont {K.~J.}\ \bibnamefont
  {Schrenk}}, \bibinfo {author} {\bibfnamefont {S.}~\bibnamefont {Rouhani}}, \
  and\ \bibinfo {author} {\bibfnamefont {H.~J.}\ \bibnamefont {Herrmann}},\
  }\href {\doibase 10.1103/PhysRevLett.109.218701} {\bibfield  {journal}
  {\bibinfo  {journal} {Phys. Rev. Lett.}\ }\textbf {\bibinfo {volume} {109}},\
  \bibinfo {pages} {218701} (\bibinfo {year} {2012})}\BibitemShut {NoStop}%
\bibitem [{\citenamefont {{Andrade Jr.}}\ \emph {et~al.}(2009)\citenamefont
  {{Andrade Jr.}}, \citenamefont {Oliveira}, \citenamefont {Moreira},\ and\
  \citenamefont {Herrmann}}]{Andrade09}%
  \BibitemOpen
  \bibfield  {author} {\bibinfo {author} {\bibfnamefont {J.~S.}\ \bibnamefont
  {{Andrade Jr.}}}, \bibinfo {author} {\bibfnamefont {E.~A.}\ \bibnamefont
  {Oliveira}}, \bibinfo {author} {\bibfnamefont {A.~A.}\ \bibnamefont
  {Moreira}}, \ and\ \bibinfo {author} {\bibfnamefont {H.~J.}\ \bibnamefont
  {Herrmann}},\ }\href {\doibase 10.1103/PhysRevLett.103.225503} {\bibfield
  {journal} {\bibinfo  {journal} {Phys. Rev. Lett.}\ }\textbf {\bibinfo
  {volume} {103}},\ \bibinfo {pages} {225503} (\bibinfo {year}
  {2009})}\BibitemShut {NoStop}%
\bibitem [{\citenamefont {{Andrade Jr.}}\ \emph {et~al.}(2011)\citenamefont
  {{Andrade Jr.}}, \citenamefont {Reis}, \citenamefont {Oliveira},
  \citenamefont {Fehr},\ and\ \citenamefont {Herrmann}}]{Andrade11}%
  \BibitemOpen
  \bibfield  {author} {\bibinfo {author} {\bibfnamefont {J.~S.}\ \bibnamefont
  {{Andrade Jr.}}}, \bibinfo {author} {\bibfnamefont {S.~D.~S.}\ \bibnamefont
  {Reis}}, \bibinfo {author} {\bibfnamefont {E.~A.}\ \bibnamefont {Oliveira}},
  \bibinfo {author} {\bibfnamefont {E.}~\bibnamefont {Fehr}}, \ and\ \bibinfo
  {author} {\bibfnamefont {H.~J.}\ \bibnamefont {Herrmann}},\ }\href {\doibase
  10.1109/MCSE.2011.16} {\bibfield  {journal} {\bibinfo  {journal} {Comput.
  Sci. Eng.}\ }\textbf {\bibinfo {volume} {13}},\ \bibinfo {pages} {74}
  (\bibinfo {year} {2011})}\BibitemShut {NoStop}%
\bibitem [{\citenamefont {Oliveira}\ \emph {et~al.}(2011)\citenamefont
  {Oliveira}, \citenamefont {Schrenk}, \citenamefont {Ara\'ujo}, \citenamefont
  {Herrmann},\ and\ \citenamefont {{Andrade Jr.}}}]{Oliveira11}%
  \BibitemOpen
  \bibfield  {author} {\bibinfo {author} {\bibfnamefont {E.~A.}\ \bibnamefont
  {Oliveira}}, \bibinfo {author} {\bibfnamefont {K.~J.}\ \bibnamefont
  {Schrenk}}, \bibinfo {author} {\bibfnamefont {N.~A.~M.}\ \bibnamefont
  {Ara\'ujo}}, \bibinfo {author} {\bibfnamefont {H.~J.}\ \bibnamefont
  {Herrmann}}, \ and\ \bibinfo {author} {\bibfnamefont {J.~S.}\ \bibnamefont
  {{Andrade Jr.}}},\ }\href {\doibase 10.1103/PhysRevE.83.046113} {\bibfield
  {journal} {\bibinfo  {journal} {Phys. Rev. E}\ }\textbf {\bibinfo {volume}
  {83}},\ \bibinfo {pages} {046113} (\bibinfo {year} {2011})}\BibitemShut
  {NoStop}%
\bibitem [{\citenamefont {Moreira}\ \emph {et~al.}(2012)\citenamefont
  {Moreira}, \citenamefont {Oliveira}, \citenamefont {Hansen}, \citenamefont
  {Ara\'ujo}, \citenamefont {Herrmann},\ and\ \citenamefont {{Andrade
  Jr.}}}]{Moreira12}%
  \BibitemOpen
  \bibfield  {author} {\bibinfo {author} {\bibfnamefont {A.~A.}\ \bibnamefont
  {Moreira}}, \bibinfo {author} {\bibfnamefont {C.~L.~N.}\ \bibnamefont
  {Oliveira}}, \bibinfo {author} {\bibfnamefont {A.}~\bibnamefont {Hansen}},
  \bibinfo {author} {\bibfnamefont {N.~A.~M.}\ \bibnamefont {Ara\'ujo}},
  \bibinfo {author} {\bibfnamefont {H.~J.}\ \bibnamefont {Herrmann}}, \ and\
  \bibinfo {author} {\bibfnamefont {J.~S.}\ \bibnamefont {{Andrade Jr.}}},\
  }\href {\doibase 10.1103/PhysRevLett.109.255701} {\bibfield  {journal}
  {\bibinfo  {journal} {Phys. Rev. Lett.}\ }\textbf {\bibinfo {volume} {109}},\
  \bibinfo {pages} {255701} (\bibinfo {year} {2012})}\BibitemShut {NoStop}%
\bibitem [{\citenamefont {Stark}(1991)}]{Stark91}%
  \BibitemOpen
  \bibfield  {author} {\bibinfo {author} {\bibfnamefont {C.~P.}\ \bibnamefont
  {Stark}},\ }\href@noop {} {\bibfield  {journal} {\bibinfo  {journal}
  {Nature}\ }\textbf {\bibinfo {volume} {352}},\ \bibinfo {pages} {423}
  (\bibinfo {year} {1991})}\BibitemShut {NoStop}%
\bibitem [{\citenamefont {Maritan}\ \emph {et~al.}(1996)\citenamefont
  {Maritan}, \citenamefont {Colaiori}, \citenamefont {Flammini}, \citenamefont
  {Cieplak},\ and\ \citenamefont {Banavar}}]{Maritan96}%
  \BibitemOpen
  \bibfield  {author} {\bibinfo {author} {\bibfnamefont {A.}~\bibnamefont
  {Maritan}}, \bibinfo {author} {\bibfnamefont {F.}~\bibnamefont {Colaiori}},
  \bibinfo {author} {\bibfnamefont {A.}~\bibnamefont {Flammini}}, \bibinfo
  {author} {\bibfnamefont {M.}~\bibnamefont {Cieplak}}, \ and\ \bibinfo
  {author} {\bibfnamefont {J.~R.}\ \bibnamefont {Banavar}},\ }\href@noop {}
  {\bibfield  {journal} {\bibinfo  {journal} {Science}\ }\textbf {\bibinfo
  {volume} {272}},\ \bibinfo {pages} {984} (\bibinfo {year}
  {1996})}\BibitemShut {NoStop}%
\bibitem [{\citenamefont {Manna}\ and\ \citenamefont
  {Subramanian}(1996)}]{Manna95}%
  \BibitemOpen
  \bibfield  {author} {\bibinfo {author} {\bibfnamefont {S.~S.}\ \bibnamefont
  {Manna}}\ and\ \bibinfo {author} {\bibfnamefont {B.}~\bibnamefont
  {Subramanian}},\ }\href {\doibase 10.1103/PhysRevLett.76.3460} {\bibfield
  {journal} {\bibinfo  {journal} {Phys. Rev. Lett.}\ }\textbf {\bibinfo
  {volume} {76}},\ \bibinfo {pages} {3460} (\bibinfo {year}
  {1996})}\BibitemShut {NoStop}%
\bibitem [{\citenamefont {Knecht}\ \emph {et~al.}(2012)\citenamefont {Knecht},
  \citenamefont {Trump}, \citenamefont {{ben-Avraham}},\ and\ \citenamefont
  {Ziff}}]{Knecht11}%
  \BibitemOpen
  \bibfield  {author} {\bibinfo {author} {\bibfnamefont {C.~L.}\ \bibnamefont
  {Knecht}}, \bibinfo {author} {\bibfnamefont {W.}~\bibnamefont {Trump}},
  \bibinfo {author} {\bibfnamefont {D.}~\bibnamefont {{ben-Avraham}}}, \ and\
  \bibinfo {author} {\bibfnamefont {R.~M.}\ \bibnamefont {Ziff}},\ }\href
  {\doibase 10.1103/PhysRevLett.108.045703} {\bibfield  {journal} {\bibinfo
  {journal} {Phys. Rev. Lett.}\ }\textbf {\bibinfo {volume} {108}},\ \bibinfo
  {pages} {045703} (\bibinfo {year} {2012})}\BibitemShut {NoStop}%
\bibitem [{\citenamefont {Baek}\ and\ \citenamefont {Kim}(2012)}]{Baek11}%
  \BibitemOpen
  \bibfield  {author} {\bibinfo {author} {\bibfnamefont {S.~K.}\ \bibnamefont
  {Baek}}\ and\ \bibinfo {author} {\bibfnamefont {B.~J.}\ \bibnamefont {Kim}},\
  }\href {\doibase 10.1103/PhysRevE.85.032103} {\bibfield  {journal} {\bibinfo
  {journal} {Phys. Rev. E}\ }\textbf {\bibinfo {volume} {85}},\ \bibinfo
  {pages} {032103} (\bibinfo {year} {2012})}\BibitemShut {NoStop}%
\bibitem [{\citenamefont {Ikedo}\ \emph {et~al.}(2007)\citenamefont {Ikedo},
  \citenamefont {Fukuoka}, \citenamefont {Hara}, \citenamefont {Fujita},
  \citenamefont {Takada}, \citenamefont {Endo},\ and\ \citenamefont
  {Morita}}]{Ikedo07}%
  \BibitemOpen
  \bibfield  {author} {\bibinfo {author} {\bibfnamefont {Y.}~\bibnamefont
  {Ikedo}}, \bibinfo {author} {\bibfnamefont {D.}~\bibnamefont {Fukuoka}},
  \bibinfo {author} {\bibfnamefont {T.}~\bibnamefont {Hara}}, \bibinfo {author}
  {\bibfnamefont {H.}~\bibnamefont {Fujita}}, \bibinfo {author} {\bibfnamefont
  {E.}~\bibnamefont {Takada}}, \bibinfo {author} {\bibfnamefont
  {T.}~\bibnamefont {Endo}}, \ and\ \bibinfo {author} {\bibfnamefont
  {T.}~\bibnamefont {Morita}},\ }\href {\doibase 10.1118/1.2795825} {\bibfield
  {journal} {\bibinfo  {journal} {Med. Phys.}\ }\textbf {\bibinfo {volume}
  {34}},\ \bibinfo {pages} {4378} (\bibinfo {year} {2007})}\BibitemShut
  {NoStop}%
\bibitem [{\citenamefont {Kerr}\ \emph {et~al.}(2006)\citenamefont {Kerr},
  \citenamefont {Neuhauser}, \citenamefont {Bohannan},\ and\ \citenamefont
  {Dean}}]{Kerr06}%
  \BibitemOpen
  \bibfield  {author} {\bibinfo {author} {\bibfnamefont {B.}~\bibnamefont
  {Kerr}}, \bibinfo {author} {\bibfnamefont {C.}~\bibnamefont {Neuhauser}},
  \bibinfo {author} {\bibfnamefont {B.~J.~M.}\ \bibnamefont {Bohannan}}, \ and\
  \bibinfo {author} {\bibfnamefont {A.~M.}\ \bibnamefont {Dean}},\ }\href@noop
  {} {\bibfield  {journal} {\bibinfo  {journal} {Nature}\ }\textbf {\bibinfo
  {volume} {442}},\ \bibinfo {pages} {75} (\bibinfo {year} {2006})}\BibitemShut
  {NoStop}%
\bibitem [{\citenamefont {Mathiesen}\ \emph {et~al.}(2011)\citenamefont
  {Mathiesen}, \citenamefont {Mitarai}, \citenamefont {Sneppen},\ and\
  \citenamefont {Trusina}}]{Mathiesen11}%
  \BibitemOpen
  \bibfield  {author} {\bibinfo {author} {\bibfnamefont {J.}~\bibnamefont
  {Mathiesen}}, \bibinfo {author} {\bibfnamefont {N.}~\bibnamefont {Mitarai}},
  \bibinfo {author} {\bibfnamefont {K.}~\bibnamefont {Sneppen}}, \ and\
  \bibinfo {author} {\bibfnamefont {A.}~\bibnamefont {Trusina}},\ }\href
  {\doibase 10.1103/PhysRevLett.107.188101} {\bibfield  {journal} {\bibinfo
  {journal} {Phys. Rev. Lett.}\ }\textbf {\bibinfo {volume} {107}},\ \bibinfo
  {pages} {188101} (\bibinfo {year} {2011})}\BibitemShut {NoStop}%
\bibitem [{\citenamefont {{Pastor-Satorras}}\ and\ \citenamefont
  {Rothman}(1998)}]{Pastor-Satorras98}%
  \BibitemOpen
  \bibfield  {author} {\bibinfo {author} {\bibfnamefont {R.}~\bibnamefont
  {{Pastor-Satorras}}}\ and\ \bibinfo {author} {\bibfnamefont {D.~H.}\
  \bibnamefont {Rothman}},\ }\href {\doibase 10.1103/PhysRevLett.80.4349}
  {\bibfield  {journal} {\bibinfo  {journal} {Phys. Rev. Lett.}\ }\textbf
  {\bibinfo {volume} {80}},\ \bibinfo {pages} {4349} (\bibinfo {year}
  {1998})}\BibitemShut {NoStop}%
\bibitem [{\citenamefont {Farr}\ \emph {et~al.}(2007)\citenamefont {Farr},
  \citenamefont {Rosen}, \citenamefont {Caro}, \citenamefont {Crippen},
  \citenamefont {Duren}, \citenamefont {Hensley}, \citenamefont {Kobrick},
  \citenamefont {Paller}, \citenamefont {Rodriguez}, \citenamefont {Roth},
  \citenamefont {Seal}, \citenamefont {Shaffer}, \citenamefont {Shimada},
  \citenamefont {Umland}, \citenamefont {Werner}, \citenamefont {Oskin},
  \citenamefont {Burbank},\ and\ \citenamefont {Alsdorf}}]{Farr07}%
  \BibitemOpen
  \bibfield  {author} {\bibinfo {author} {\bibfnamefont {T.~G.}\ \bibnamefont
  {Farr}}, \bibinfo {author} {\bibfnamefont {P.~A.}\ \bibnamefont {Rosen}},
  \bibinfo {author} {\bibfnamefont {E.}~\bibnamefont {Caro}}, \bibinfo {author}
  {\bibfnamefont {R.}~\bibnamefont {Crippen}}, \bibinfo {author} {\bibfnamefont
  {R.}~\bibnamefont {Duren}}, \bibinfo {author} {\bibfnamefont
  {S.}~\bibnamefont {Hensley}}, \bibinfo {author} {\bibfnamefont
  {M.}~\bibnamefont {Kobrick}}, \bibinfo {author} {\bibfnamefont
  {M.}~\bibnamefont {Paller}}, \bibinfo {author} {\bibfnamefont
  {E.}~\bibnamefont {Rodriguez}}, \bibinfo {author} {\bibfnamefont
  {L.}~\bibnamefont {Roth}}, \bibinfo {author} {\bibfnamefont {D.}~\bibnamefont
  {Seal}}, \bibinfo {author} {\bibfnamefont {S.}~\bibnamefont {Shaffer}},
  \bibinfo {author} {\bibfnamefont {J.}~\bibnamefont {Shimada}}, \bibinfo
  {author} {\bibfnamefont {J.}~\bibnamefont {Umland}}, \bibinfo {author}
  {\bibfnamefont {M.}~\bibnamefont {Werner}}, \bibinfo {author} {\bibfnamefont
  {M.}~\bibnamefont {Oskin}}, \bibinfo {author} {\bibfnamefont
  {D.}~\bibnamefont {Burbank}}, \ and\ \bibinfo {author} {\bibfnamefont
  {D.}~\bibnamefont {Alsdorf}},\ }\href {\doibase 10.1029/2005RG000183}
  {\bibfield  {journal} {\bibinfo  {journal} {Rev. Geophys.}\ }\textbf
  {\bibinfo {volume} {45}},\ \bibinfo {pages} {RG2004} (\bibinfo {year}
  {2007})}\BibitemShut {NoStop}%
\bibitem [{\citenamefont {Vincent}\ and\ \citenamefont
  {Soille}(1991)}]{Vincent91}%
  \BibitemOpen
  \bibfield  {author} {\bibinfo {author} {\bibfnamefont {L.}~\bibnamefont
  {Vincent}}\ and\ \bibinfo {author} {\bibfnamefont {P.}~\bibnamefont
  {Soille}},\ }\href@noop {} {\bibfield  {journal} {\bibinfo  {journal} {IEEE
  T. Pattern Anal.}\ }\textbf {\bibinfo {volume} {13}},\ \bibinfo {pages} {583}
  (\bibinfo {year} {1991})}\BibitemShut {NoStop}%
\bibitem [{\citenamefont {Fehr}\ \emph {et~al.}(2009)\citenamefont {Fehr},
  \citenamefont {{Andrade Jr.}}, \citenamefont {{da Cunha}}, \citenamefont {{da
  Silva}}, \citenamefont {Herrmann}, \citenamefont {Kadau}, \citenamefont
  {Moukarzel},\ and\ \citenamefont {Oliveira}}]{Fehr09}%
  \BibitemOpen
  \bibfield  {author} {\bibinfo {author} {\bibfnamefont {E.}~\bibnamefont
  {Fehr}}, \bibinfo {author} {\bibfnamefont {J.~S.}\ \bibnamefont {{Andrade
  Jr.}}}, \bibinfo {author} {\bibfnamefont {S.~D.}\ \bibnamefont {{da Cunha}}},
  \bibinfo {author} {\bibfnamefont {L.~R.}\ \bibnamefont {{da Silva}}},
  \bibinfo {author} {\bibfnamefont {H.~J.}\ \bibnamefont {Herrmann}}, \bibinfo
  {author} {\bibfnamefont {D.}~\bibnamefont {Kadau}}, \bibinfo {author}
  {\bibfnamefont {C.~F.}\ \bibnamefont {Moukarzel}}, \ and\ \bibinfo {author}
  {\bibfnamefont {E.~A.}\ \bibnamefont {Oliveira}},\ }\href {\doibase
  10.1088/1742-5468/2009/09/P09007} {\bibfield  {journal} {\bibinfo  {journal}
  {J. Stat. Mech.}\ ,\ \bibinfo {pages} {P09007}} (\bibinfo {year}
  {2009})}\BibitemShut {NoStop}%
\bibitem [{\citenamefont {Wilkinson}\ and\ \citenamefont
  {Willemsen}(1983)}]{Wilkinson83}%
  \BibitemOpen
  \bibfield  {author} {\bibinfo {author} {\bibfnamefont {D.}~\bibnamefont
  {Wilkinson}}\ and\ \bibinfo {author} {\bibfnamefont {J.~F.}\ \bibnamefont
  {Willemsen}},\ }\href {\doibase 10.1088/0305-4470/16/14/028} {\bibfield
  {journal} {\bibinfo  {journal} {J. Phys. A}\ }\textbf {\bibinfo {volume}
  {16}},\ \bibinfo {pages} {3365} (\bibinfo {year} {1983})}\BibitemShut
  {NoStop}%
\bibitem [{\citenamefont {Fehr}\ \emph
  {et~al.}(2011{\natexlab{a}})\citenamefont {Fehr}, \citenamefont {Kadau},
  \citenamefont {Ara\'ujo}, \citenamefont {{Andrade Jr.}},\ and\ \citenamefont
  {Herrmann}}]{Fehr11b}%
  \BibitemOpen
  \bibfield  {author} {\bibinfo {author} {\bibfnamefont {E.}~\bibnamefont
  {Fehr}}, \bibinfo {author} {\bibfnamefont {D.}~\bibnamefont {Kadau}},
  \bibinfo {author} {\bibfnamefont {N.~A.~M.}\ \bibnamefont {Ara\'ujo}},
  \bibinfo {author} {\bibfnamefont {J.~S.}\ \bibnamefont {{Andrade Jr.}}}, \
  and\ \bibinfo {author} {\bibfnamefont {H.~J.}\ \bibnamefont {Herrmann}},\
  }\href {\doibase 10.1103/PhysRevE.84.036116} {\bibfield  {journal} {\bibinfo
  {journal} {Phys. Rev. E}\ }\textbf {\bibinfo {volume} {84}},\ \bibinfo
  {pages} {036116} (\bibinfo {year} {2011}{\natexlab{a}})}\BibitemShut
  {NoStop}%
\bibitem [{\citenamefont {Breyer}\ and\ \citenamefont {Snow}(1992)}]{Breyer92}%
  \BibitemOpen
  \bibfield  {author} {\bibinfo {author} {\bibfnamefont {S.~P.}\ \bibnamefont
  {Breyer}}\ and\ \bibinfo {author} {\bibfnamefont {R.~S.}\ \bibnamefont
  {Snow}},\ }\href@noop {} {\bibfield  {journal} {\bibinfo  {journal}
  {Geomorphology}\ }\textbf {\bibinfo {volume} {5}},\ \bibinfo {pages} {143}
  (\bibinfo {year} {1992})}\BibitemShut {NoStop}%
\bibitem [{\citenamefont {Mandelbrot}(1967)}]{Mandelbrot67}%
  \BibitemOpen
  \bibfield  {author} {\bibinfo {author} {\bibfnamefont {B.}~\bibnamefont
  {Mandelbrot}},\ }\href@noop {} {\bibfield  {journal} {\bibinfo  {journal}
  {Science}\ }\textbf {\bibinfo {volume} {156}},\ \bibinfo {pages} {636}
  (\bibinfo {year} {1967})}\BibitemShut {NoStop}%
\bibitem [{\citenamefont {Mandelbrot}(1983)}]{Mandelbrot83}%
  \BibitemOpen
  \bibfield  {author} {\bibinfo {author} {\bibfnamefont {B.~B.}\ \bibnamefont
  {Mandelbrot}},\ }\href@noop {} {\emph {\bibinfo {title} {The Fractal Geometry
  of Nature}}}\ (\bibinfo  {publisher} {Freeman},\ \bibinfo {address} {New
  York},\ \bibinfo {year} {1983})\BibitemShut {NoStop}%
\bibitem [{\citenamefont {Cieplak}\ \emph {et~al.}(1994)\citenamefont
  {Cieplak}, \citenamefont {Maritan},\ and\ \citenamefont
  {Banavar}}]{Cieplak94}%
  \BibitemOpen
  \bibfield  {author} {\bibinfo {author} {\bibfnamefont {M.}~\bibnamefont
  {Cieplak}}, \bibinfo {author} {\bibfnamefont {A.}~\bibnamefont {Maritan}}, \
  and\ \bibinfo {author} {\bibfnamefont {J.~R.}\ \bibnamefont {Banavar}},\
  }\href {\doibase 10.1103/PhysRevLett.72.2320} {\bibfield  {journal} {\bibinfo
   {journal} {Phys. Rev. Lett.}\ }\textbf {\bibinfo {volume} {72}},\ \bibinfo
  {pages} {2320} (\bibinfo {year} {1994})}\BibitemShut {NoStop}%
\bibitem [{\citenamefont {Fehr}\ \emph {et~al.}(2012)\citenamefont {Fehr},
  \citenamefont {Schrenk}, \citenamefont {Ara\'ujo}, \citenamefont {Kadau},
  \citenamefont {Grassberger}, \citenamefont {{Andrade Jr.}},\ and\
  \citenamefont {Herrmann}}]{Fehr11c}%
  \BibitemOpen
  \bibfield  {author} {\bibinfo {author} {\bibfnamefont {E.}~\bibnamefont
  {Fehr}}, \bibinfo {author} {\bibfnamefont {K.~J.}\ \bibnamefont {Schrenk}},
  \bibinfo {author} {\bibfnamefont {N.~A.~M.}\ \bibnamefont {Ara\'ujo}},
  \bibinfo {author} {\bibfnamefont {D.}~\bibnamefont {Kadau}}, \bibinfo
  {author} {\bibfnamefont {P.}~\bibnamefont {Grassberger}}, \bibinfo {author}
  {\bibfnamefont {J.~S.}\ \bibnamefont {{Andrade Jr.}}}, \ and\ \bibinfo
  {author} {\bibfnamefont {H.~J.}\ \bibnamefont {Herrmann}},\ }\href {\doibase
  10.1103/PhysRevE.86.011117} {\bibfield  {journal} {\bibinfo  {journal} {Phys.
  Rev. E}\ }\textbf {\bibinfo {volume} {86}},\ \bibinfo {pages} {011117}
  (\bibinfo {year} {2012})}\BibitemShut {NoStop}%
\bibitem [{\citenamefont {Porto}\ \emph {et~al.}(1997)\citenamefont {Porto},
  \citenamefont {Havlin}, \citenamefont {Schwarzer},\ and\ \citenamefont
  {Bunde}}]{Porto97}%
  \BibitemOpen
  \bibfield  {author} {\bibinfo {author} {\bibfnamefont {M.}~\bibnamefont
  {Porto}}, \bibinfo {author} {\bibfnamefont {S.}~\bibnamefont {Havlin}},
  \bibinfo {author} {\bibfnamefont {S.}~\bibnamefont {Schwarzer}}, \ and\
  \bibinfo {author} {\bibfnamefont {A.}~\bibnamefont {Bunde}},\ }\href
  {\doibase 10.1103/PhysRevLett.79.4060} {\bibfield  {journal} {\bibinfo
  {journal} {Phys. Rev. Lett.}\ }\textbf {\bibinfo {volume} {79}},\ \bibinfo
  {pages} {4060} (\bibinfo {year} {1997})}\BibitemShut {NoStop}%
\bibitem [{\citenamefont {Buldyrev}\ \emph {et~al.}(2004)\citenamefont
  {Buldyrev}, \citenamefont {Havlin}, \citenamefont {L{\'o}pez},\ and\
  \citenamefont {Stanley}}]{Buldyrev04}%
  \BibitemOpen
  \bibfield  {author} {\bibinfo {author} {\bibfnamefont {S.~V.}\ \bibnamefont
  {Buldyrev}}, \bibinfo {author} {\bibfnamefont {S.}~\bibnamefont {Havlin}},
  \bibinfo {author} {\bibfnamefont {E.}~\bibnamefont {L{\'o}pez}}, \ and\
  \bibinfo {author} {\bibfnamefont {H.~E.}\ \bibnamefont {Stanley}},\ }\href
  {\doibase 10.1103/PhysRevE.70.035102} {\bibfield  {journal} {\bibinfo
  {journal} {Phys. Rev. E}\ }\textbf {\bibinfo {volume} {70}},\ \bibinfo
  {pages} {035102(R)} (\bibinfo {year} {2004})}\BibitemShut {NoStop}%
\bibitem [{\citenamefont {Cieplak}\ \emph {et~al.}(1996)\citenamefont
  {Cieplak}, \citenamefont {Maritan},\ and\ \citenamefont
  {Banavar}}]{Cieplak96}%
  \BibitemOpen
  \bibfield  {author} {\bibinfo {author} {\bibfnamefont {M.}~\bibnamefont
  {Cieplak}}, \bibinfo {author} {\bibfnamefont {A.}~\bibnamefont {Maritan}}, \
  and\ \bibinfo {author} {\bibfnamefont {J.~R.}\ \bibnamefont {Banavar}},\
  }\href {\doibase 10.1103/PhysRevLett.76.3754} {\bibfield  {journal} {\bibinfo
   {journal} {Phys. Rev. Lett.}\ }\textbf {\bibinfo {volume} {76}},\ \bibinfo
  {pages} {3754} (\bibinfo {year} {1996})}\BibitemShut {NoStop}%
\bibitem [{\citenamefont {Ara\'ujo}\ and\ \citenamefont
  {Herrmann}(2010)}]{Araujo10}%
  \BibitemOpen
  \bibfield  {author} {\bibinfo {author} {\bibfnamefont {N.~A.~M.}\
  \bibnamefont {Ara\'ujo}}\ and\ \bibinfo {author} {\bibfnamefont {H.~J.}\
  \bibnamefont {Herrmann}},\ }\href {\doibase 10.1103/PhysRevLett.105.035701}
  {\bibfield  {journal} {\bibinfo  {journal} {Phys. Rev. Lett.}\ }\textbf
  {\bibinfo {volume} {105}},\ \bibinfo {pages} {035701} (\bibinfo {year}
  {2010})}\BibitemShut {NoStop}%
\bibitem [{\citenamefont {Schrenk}\ \emph {et~al.}(2011)\citenamefont
  {Schrenk}, \citenamefont {Ara\'ujo},\ and\ \citenamefont
  {Herrmann}}]{Schrenk11}%
  \BibitemOpen
  \bibfield  {author} {\bibinfo {author} {\bibfnamefont {K.~J.}\ \bibnamefont
  {Schrenk}}, \bibinfo {author} {\bibfnamefont {N.~A.~M.}\ \bibnamefont
  {Ara\'ujo}}, \ and\ \bibinfo {author} {\bibfnamefont {H.~J.}\ \bibnamefont
  {Herrmann}},\ }\href {\doibase 10.1103/PhysRevE.84.041136} {\bibfield
  {journal} {\bibinfo  {journal} {Phys. Rev. E}\ }\textbf {\bibinfo {volume}
  {84}},\ \bibinfo {pages} {041136} (\bibinfo {year} {2011})}\BibitemShut
  {NoStop}%
\bibitem [{\citenamefont {Douglass}\ and\ \citenamefont
  {Schmeeckle}(2007)}]{Douglass07}%
  \BibitemOpen
  \bibfield  {author} {\bibinfo {author} {\bibfnamefont {J.}~\bibnamefont
  {Douglass}}\ and\ \bibinfo {author} {\bibfnamefont {M.}~\bibnamefont
  {Schmeeckle}},\ }\href {\doibase 10.1016/j.geomorph.2006.06.004} {\bibfield
  {journal} {\bibinfo  {journal} {Geomorphology}\ }\textbf {\bibinfo {volume}
  {84}},\ \bibinfo {pages} {22} (\bibinfo {year} {2007})}\BibitemShut {NoStop}%
\bibitem [{\citenamefont {Bishop}(1995)}]{Bishop95}%
  \BibitemOpen
  \bibfield  {author} {\bibinfo {author} {\bibfnamefont {P.}~\bibnamefont
  {Bishop}},\ }\href {\doibase 10.1177/030913339501900402} {\bibfield
  {journal} {\bibinfo  {journal} {Prog. Phys. Geog.}\ }\textbf {\bibinfo
  {volume} {19}},\ \bibinfo {pages} {449} (\bibinfo {year} {1995})}\BibitemShut
  {NoStop}%
\bibitem [{\citenamefont {Burridge}\ \emph {et~al.}(2007)\citenamefont
  {Burridge}, \citenamefont {Craw},\ and\ \citenamefont {Waters}}]{Burridge07}%
  \BibitemOpen
  \bibfield  {author} {\bibinfo {author} {\bibfnamefont {C.~P.}\ \bibnamefont
  {Burridge}}, \bibinfo {author} {\bibfnamefont {D.}~\bibnamefont {Craw}}, \
  and\ \bibinfo {author} {\bibfnamefont {J.~M.}\ \bibnamefont {Waters}},\
  }\href {\doibase 10.1111/j.1365-294X.2006.03196.x} {\bibfield  {journal}
  {\bibinfo  {journal} {Mol. Ecol.}\ }\textbf {\bibinfo {volume} {16}},\
  \bibinfo {pages} {1883} (\bibinfo {year} {2007})}\BibitemShut {NoStop}%
\bibitem [{\citenamefont {{Garcia-Castellanos}}\ \emph
  {et~al.}(2009)\citenamefont {{Garcia-Castellanos}}, \citenamefont {Estrada},
  \citenamefont {{Jim{\'e}nez-Munt}}, \citenamefont {Gorini}, \citenamefont
  {Fern{\`a}ndez}, \citenamefont {Verg{\'e}s},\ and\ \citenamefont {{De
  Vicente}}}]{Garcia-Castellanos09}%
  \BibitemOpen
  \bibfield  {author} {\bibinfo {author} {\bibfnamefont {D.}~\bibnamefont
  {{Garcia-Castellanos}}}, \bibinfo {author} {\bibfnamefont {F.}~\bibnamefont
  {Estrada}}, \bibinfo {author} {\bibfnamefont {I.}~\bibnamefont
  {{Jim{\'e}nez-Munt}}}, \bibinfo {author} {\bibfnamefont {C.}~\bibnamefont
  {Gorini}}, \bibinfo {author} {\bibfnamefont {M.}~\bibnamefont
  {Fern{\`a}ndez}}, \bibinfo {author} {\bibfnamefont {J.}~\bibnamefont
  {Verg{\'e}s}}, \ and\ \bibinfo {author} {\bibfnamefont {R.}~\bibnamefont {{De
  Vicente}}},\ }\href {\doibase 10.1038/nature08555} {\bibfield  {journal}
  {\bibinfo  {journal} {Nature}\ }\textbf {\bibinfo {volume} {462}},\ \bibinfo
  {pages} {778} (\bibinfo {year} {2009})}\BibitemShut {NoStop}%
\bibitem [{\citenamefont {Linkevi{\v c}ien{\. e}}(2009)}]{Linkeviciene09}%
  \BibitemOpen
  \bibfield  {author} {\bibinfo {author} {\bibfnamefont {R.}~\bibnamefont
  {Linkevi{\v c}ien{\. e}}},\ }\href {\doibase 10.1177/0959683609345081}
  {\bibfield  {journal} {\bibinfo  {journal} {Holocene}\ }\textbf {\bibinfo
  {volume} {19}},\ \bibinfo {pages} {1233} (\bibinfo {year}
  {2009})}\BibitemShut {NoStop}%
\bibitem [{\citenamefont {{Garcia-Castellanos}}\ \emph
  {et~al.}(2003)\citenamefont {{Garcia-Castellanos}}, \citenamefont
  {Verg{\'e}s}, \citenamefont {{Gaspar-Escribano}},\ and\ \citenamefont
  {Cloetingh}}]{Garcia-Castellanos03}%
  \BibitemOpen
  \bibfield  {author} {\bibinfo {author} {\bibfnamefont {D.}~\bibnamefont
  {{Garcia-Castellanos}}}, \bibinfo {author} {\bibfnamefont {J.}~\bibnamefont
  {Verg{\'e}s}}, \bibinfo {author} {\bibfnamefont {J.}~\bibnamefont
  {{Gaspar-Escribano}}}, \ and\ \bibinfo {author} {\bibfnamefont
  {S.}~\bibnamefont {Cloetingh}},\ }\href {\doibase 10.1029/2002JB002073}
  {\bibfield  {journal} {\bibinfo  {journal} {J. Geophys. Res.}\ }\textbf
  {\bibinfo {volume} {108}},\ \bibinfo {pages} {2347} (\bibinfo {year}
  {2003})}\BibitemShut {NoStop}%
\bibitem [{\citenamefont {Dorsey}\ and\ \citenamefont
  {Roering}(2006)}]{Dorsey06}%
  \BibitemOpen
  \bibfield  {author} {\bibinfo {author} {\bibfnamefont {R.~J.}\ \bibnamefont
  {Dorsey}}\ and\ \bibinfo {author} {\bibfnamefont {J.~J.}\ \bibnamefont
  {Roering}},\ }\href {\doibase 10.1016/j.geomorph.2005.06.013} {\bibfield
  {journal} {\bibinfo  {journal} {Geomorphology}\ }\textbf {\bibinfo {volume}
  {73}},\ \bibinfo {pages} {16} (\bibinfo {year} {2006})}\BibitemShut {NoStop}%
\bibitem [{\citenamefont {Lock}\ \emph {et~al.}(2006)\citenamefont {Lock},
  \citenamefont {Kelsey}, \citenamefont {Furlong},\ and\ \citenamefont
  {Woolace}}]{Lock06}%
  \BibitemOpen
  \bibfield  {author} {\bibinfo {author} {\bibfnamefont {J.}~\bibnamefont
  {Lock}}, \bibinfo {author} {\bibfnamefont {H.}~\bibnamefont {Kelsey}},
  \bibinfo {author} {\bibfnamefont {K.}~\bibnamefont {Furlong}}, \ and\
  \bibinfo {author} {\bibfnamefont {A.}~\bibnamefont {Woolace}},\ }\href
  {\doibase 10.1130/B25885.1} {\bibfield  {journal} {\bibinfo  {journal} {Geol.
  Soc. Am. Bull.}\ }\textbf {\bibinfo {volume} {118}},\ \bibinfo {pages} {1232}
  (\bibinfo {year} {2006})}\BibitemShut {NoStop}%
\bibitem [{\citenamefont {Beranek}\ \emph {et~al.}(2006)\citenamefont
  {Beranek}, \citenamefont {Link},\ and\ \citenamefont {Fanning}}]{Beranek06}%
  \BibitemOpen
  \bibfield  {author} {\bibinfo {author} {\bibfnamefont {L.~P.}\ \bibnamefont
  {Beranek}}, \bibinfo {author} {\bibfnamefont {P.~K.}\ \bibnamefont {Link}}, \
  and\ \bibinfo {author} {\bibfnamefont {C.~M.}\ \bibnamefont {Fanning}},\
  }\href {\doibase 10.1130/B25896.1} {\bibfield  {journal} {\bibinfo  {journal}
  {Geol. Soc. Am. Bull.}\ }\textbf {\bibinfo {volume} {118}},\ \bibinfo {pages}
  {1027} (\bibinfo {year} {2006})}\BibitemShut {NoStop}%
\bibitem [{\citenamefont {Attal}\ \emph {et~al.}(2008)\citenamefont {Attal},
  \citenamefont {Tucker}, \citenamefont {Whittaker}, \citenamefont {Cowie},\
  and\ \citenamefont {Roberts}}]{Attal08}%
  \BibitemOpen
  \bibfield  {author} {\bibinfo {author} {\bibfnamefont {M.}~\bibnamefont
  {Attal}}, \bibinfo {author} {\bibfnamefont {G.~E.}\ \bibnamefont {Tucker}},
  \bibinfo {author} {\bibfnamefont {A.~C.}\ \bibnamefont {Whittaker}}, \bibinfo
  {author} {\bibfnamefont {P.~A.}\ \bibnamefont {Cowie}}, \ and\ \bibinfo
  {author} {\bibfnamefont {G.~P.}\ \bibnamefont {Roberts}},\ }\href {\doibase
  10.1029/2007JF000893} {\bibfield  {journal} {\bibinfo  {journal} {J. Geophys.
  Res.}\ }\textbf {\bibinfo {volume} {113}},\ \bibinfo {pages} {F03013}
  (\bibinfo {year} {2008})}\BibitemShut {NoStop}%
\bibitem [{\citenamefont {Fehr}\ \emph
  {et~al.}(2011{\natexlab{b}})\citenamefont {Fehr}, \citenamefont {Kadau},
  \citenamefont {{Andrade Jr.}},\ and\ \citenamefont {Herrmann}}]{Fehr11}%
  \BibitemOpen
  \bibfield  {author} {\bibinfo {author} {\bibfnamefont {E.}~\bibnamefont
  {Fehr}}, \bibinfo {author} {\bibfnamefont {D.}~\bibnamefont {Kadau}},
  \bibinfo {author} {\bibfnamefont {J.~S.}\ \bibnamefont {{Andrade Jr.}}}, \
  and\ \bibinfo {author} {\bibfnamefont {H.~J.}\ \bibnamefont {Herrmann}},\
  }\href {\doibase 10.1103/PhysRevLett.106.048501} {\bibfield  {journal}
  {\bibinfo  {journal} {Phys. Rev. Lett.}\ }\textbf {\bibinfo {volume} {106}},\
  \bibinfo {pages} {048501} (\bibinfo {year} {2011}{\natexlab{b}})}\BibitemShut
  {NoStop}%
\bibitem [{\citenamefont {Stauffer}\ and\ \citenamefont
  {Aharony}(1994)}]{Stauffer94}%
  \BibitemOpen
  \bibfield  {author} {\bibinfo {author} {\bibfnamefont {D.}~\bibnamefont
  {Stauffer}}\ and\ \bibinfo {author} {\bibfnamefont {A.}~\bibnamefont
  {Aharony}},\ }\href@noop {} {\emph {\bibinfo {title} {Introduction to
  Percolation Theory}}},\ \bibinfo {edition} {2nd}\ ed.\ (\bibinfo  {publisher}
  {Taylor and Francis},\ \bibinfo {address} {London},\ \bibinfo {year}
  {1994})\BibitemShut {NoStop}%
\bibitem [{\citenamefont {Ara{\'u}jo}\ \emph {et~al.}(2005)\citenamefont
  {Ara{\'u}jo}, \citenamefont {Vasconcelos}, \citenamefont {Moreira},
  \citenamefont {Lucena},\ and\ \citenamefont {{Andrade Jr.}}}]{Araujo05}%
  \BibitemOpen
  \bibfield  {author} {\bibinfo {author} {\bibfnamefont {A.~D.}\ \bibnamefont
  {Ara{\'u}jo}}, \bibinfo {author} {\bibfnamefont {T.~F.}\ \bibnamefont
  {Vasconcelos}}, \bibinfo {author} {\bibfnamefont {A.~A.}\ \bibnamefont
  {Moreira}}, \bibinfo {author} {\bibfnamefont {L.~S.}\ \bibnamefont {Lucena}},
  \ and\ \bibinfo {author} {\bibfnamefont {J.~S.}\ \bibnamefont {{Andrade
  Jr.}}},\ }\href {\doibase 10.1103/PhysRevE.72.041404} {\bibfield  {journal}
  {\bibinfo  {journal} {Phys. Rev. E}\ }\textbf {\bibinfo {volume} {72}},\
  \bibinfo {pages} {041404} (\bibinfo {year} {2005})}\BibitemShut {NoStop}%
\bibitem [{\citenamefont {Lee}(2009)}]{Lee09}%
  \BibitemOpen
  \bibfield  {author} {\bibinfo {author} {\bibfnamefont {S.~B.}\ \bibnamefont
  {Lee}},\ }\href {\doibase 10.1016/j.physa.2009.03.002} {\bibfield  {journal}
  {\bibinfo  {journal} {Physica A}\ }\textbf {\bibinfo {volume} {388}},\
  \bibinfo {pages} {2271} (\bibinfo {year} {2009})}\BibitemShut {NoStop}%
\bibitem [{\citenamefont {Peitgen}\ and\ \citenamefont
  {Saupe}(1988)}]{Peitgen88}%
  \BibitemOpen
  \bibinfo {editor} {\bibfnamefont {H.-O.}\ \bibnamefont {Peitgen}}\ and\
  \bibinfo {editor} {\bibfnamefont {D.}~\bibnamefont {Saupe}},\ eds.,\
  \href@noop {} {\emph {\bibinfo {title} {The Science of Fractal Images}}}\
  (\bibinfo  {publisher} {Springer},\ \bibinfo {address} {New York},\ \bibinfo
  {year} {1988})\BibitemShut {NoStop}%
\bibitem [{\citenamefont {Morais}\ \emph {et~al.}(2011)\citenamefont {Morais},
  \citenamefont {Oliveira}, \citenamefont {Ara\'ujo}, \citenamefont
  {Herrmann},\ and\ \citenamefont {{Andrade Jr.}}}]{Morais11}%
  \BibitemOpen
  \bibfield  {author} {\bibinfo {author} {\bibfnamefont {P.~A.}\ \bibnamefont
  {Morais}}, \bibinfo {author} {\bibfnamefont {E.~A.}\ \bibnamefont
  {Oliveira}}, \bibinfo {author} {\bibfnamefont {N.~A.~M.}\ \bibnamefont
  {Ara\'ujo}}, \bibinfo {author} {\bibfnamefont {H.~J.}\ \bibnamefont
  {Herrmann}}, \ and\ \bibinfo {author} {\bibfnamefont {J.~S.}\ \bibnamefont
  {{Andrade Jr.}}},\ }\href {\doibase 10.1103/PhysRevE.84.016102} {\bibfield
  {journal} {\bibinfo  {journal} {Phys. Rev. E}\ }\textbf {\bibinfo {volume}
  {84}},\ \bibinfo {pages} {016102} (\bibinfo {year} {2011})}\BibitemShut
  {NoStop}%
\bibitem [{\citenamefont {Sahimi}(1994)}]{Sahimi94a}%
  \BibitemOpen
  \bibfield  {author} {\bibinfo {author} {\bibfnamefont {M.}~\bibnamefont
  {Sahimi}},\ }\href {\doibase 10.1051/jp1:1994107} {\bibfield  {journal}
  {\bibinfo  {journal} {J. Phys. I France}\ }\textbf {\bibinfo {volume} {4}},\
  \bibinfo {pages} {1263} (\bibinfo {year} {1994})}\BibitemShut {NoStop}%
\bibitem [{\citenamefont {Sahimi}\ and\ \citenamefont
  {Mukhopadhyay}(1996)}]{Sahimi96}%
  \BibitemOpen
  \bibfield  {author} {\bibinfo {author} {\bibfnamefont {M.}~\bibnamefont
  {Sahimi}}\ and\ \bibinfo {author} {\bibfnamefont {S.}~\bibnamefont
  {Mukhopadhyay}},\ }\href {\doibase 10.1103/PhysRevE.54.3870} {\bibfield
  {journal} {\bibinfo  {journal} {Phys. Rev. E}\ }\textbf {\bibinfo {volume}
  {54}},\ \bibinfo {pages} {3870} (\bibinfo {year} {1996})}\BibitemShut
  {NoStop}%
\bibitem [{\citenamefont {Makse}\ \emph {et~al.}(1996)\citenamefont {Makse},
  \citenamefont {Havlin}, \citenamefont {Schwartz},\ and\ \citenamefont
  {Stanley}}]{Makse96}%
  \BibitemOpen
  \bibfield  {author} {\bibinfo {author} {\bibfnamefont {H.~A.}\ \bibnamefont
  {Makse}}, \bibinfo {author} {\bibfnamefont {S.}~\bibnamefont {Havlin}},
  \bibinfo {author} {\bibfnamefont {M.}~\bibnamefont {Schwartz}}, \ and\
  \bibinfo {author} {\bibfnamefont {H.~E.}\ \bibnamefont {Stanley}},\ }\href
  {\doibase 10.1103/PhysRevE.53.5445} {\bibfield  {journal} {\bibinfo
  {journal} {Phys. Rev. E}\ }\textbf {\bibinfo {volume} {53}},\ \bibinfo
  {pages} {5445} (\bibinfo {year} {1996})}\BibitemShut {NoStop}%
\bibitem [{\citenamefont {Prakash}\ \emph {et~al.}(1992)\citenamefont
  {Prakash}, \citenamefont {Havlin}, \citenamefont {Schwartz},\ and\
  \citenamefont {Stanley}}]{Prakash92}%
  \BibitemOpen
  \bibfield  {author} {\bibinfo {author} {\bibfnamefont {S.}~\bibnamefont
  {Prakash}}, \bibinfo {author} {\bibfnamefont {S.}~\bibnamefont {Havlin}},
  \bibinfo {author} {\bibfnamefont {M.}~\bibnamefont {Schwartz}}, \ and\
  \bibinfo {author} {\bibfnamefont {H.~E.}\ \bibnamefont {Stanley}},\ }\href
  {\doibase 10.1103/PhysRevA.46.R1724} {\bibfield  {journal} {\bibinfo
  {journal} {Phys. Rev. A}\ }\textbf {\bibinfo {volume} {46}},\ \bibinfo
  {pages} {R1724} (\bibinfo {year} {1992})}\BibitemShut {NoStop}%
\bibitem [{\citenamefont {Kikkinides}\ and\ \citenamefont
  {Burganos}(1999)}]{Kikkinides99}%
  \BibitemOpen
  \bibfield  {author} {\bibinfo {author} {\bibfnamefont {E.~S.}\ \bibnamefont
  {Kikkinides}}\ and\ \bibinfo {author} {\bibfnamefont {V.~N.}\ \bibnamefont
  {Burganos}},\ }\href {\doibase 10.1103/PhysRevE.59.7185} {\bibfield
  {journal} {\bibinfo  {journal} {Phys. Rev. E}\ }\textbf {\bibinfo {volume}
  {59}},\ \bibinfo {pages} {7185} (\bibinfo {year} {1999})}\BibitemShut
  {NoStop}%
\bibitem [{\citenamefont {Stanley}\ \emph {et~al.}(1999)\citenamefont
  {Stanley}, \citenamefont {{Andrade Jr.}}, \citenamefont {Havlin},
  \citenamefont {Makse},\ and\ \citenamefont {Suki}}]{Stanley99}%
  \BibitemOpen
  \bibfield  {author} {\bibinfo {author} {\bibfnamefont {H.~E.}\ \bibnamefont
  {Stanley}}, \bibinfo {author} {\bibfnamefont {J.~S.}\ \bibnamefont {{Andrade
  Jr.}}}, \bibinfo {author} {\bibfnamefont {S.}~\bibnamefont {Havlin}},
  \bibinfo {author} {\bibfnamefont {H.~A.}\ \bibnamefont {Makse}}, \ and\
  \bibinfo {author} {\bibfnamefont {B.}~\bibnamefont {Suki}},\ }\href {\doibase
  10.1016/S0378-4371(99)00029-1} {\bibfield  {journal} {\bibinfo  {journal}
  {Physica A}\ }\textbf {\bibinfo {volume} {266}},\ \bibinfo {pages} {5}
  (\bibinfo {year} {1999})}\BibitemShut {NoStop}%
\bibitem [{\citenamefont {Makse}\ \emph {et~al.}(2000)\citenamefont {Makse},
  \citenamefont {{Andrade Jr.}},\ and\ \citenamefont {Stanley}}]{Makse00}%
  \BibitemOpen
  \bibfield  {author} {\bibinfo {author} {\bibfnamefont {H.~A.}\ \bibnamefont
  {Makse}}, \bibinfo {author} {\bibfnamefont {J.~S.}\ \bibnamefont {{Andrade
  Jr.}}}, \ and\ \bibinfo {author} {\bibfnamefont {H.~E.}\ \bibnamefont
  {Stanley}},\ }\href {\doibase 10.1103/PhysRevE.61.583} {\bibfield  {journal}
  {\bibinfo  {journal} {Phys. Rev. E}\ }\textbf {\bibinfo {volume} {61}},\
  \bibinfo {pages} {583} (\bibinfo {year} {2000})}\BibitemShut {NoStop}%
\bibitem [{\citenamefont {Ara\'ujo}\ \emph {et~al.}(2002)\citenamefont
  {Ara\'ujo}, \citenamefont {Moreira}, \citenamefont {Makse}, \citenamefont
  {Stanley},\ and\ \citenamefont {{Andrade Jr.}}}]{Araujo02}%
  \BibitemOpen
  \bibfield  {author} {\bibinfo {author} {\bibfnamefont {A.~D.}\ \bibnamefont
  {Ara\'ujo}}, \bibinfo {author} {\bibfnamefont {A.~A.}\ \bibnamefont
  {Moreira}}, \bibinfo {author} {\bibfnamefont {H.~A.}\ \bibnamefont {Makse}},
  \bibinfo {author} {\bibfnamefont {H.~E.}\ \bibnamefont {Stanley}}, \ and\
  \bibinfo {author} {\bibfnamefont {J.~S.}\ \bibnamefont {{Andrade Jr.}}},\
  }\href {\doibase 10.1103/PhysRevE.66.046304} {\bibfield  {journal} {\bibinfo
  {journal} {Phys. Rev. E}\ }\textbf {\bibinfo {volume} {66}},\ \bibinfo
  {pages} {046304} (\bibinfo {year} {2002})}\BibitemShut {NoStop}%
\bibitem [{\citenamefont {Ara\'ujo}\ \emph {et~al.}(2003)\citenamefont
  {Ara\'ujo}, \citenamefont {Moreira}, \citenamefont {{Costa Filho}},\ and\
  \citenamefont {{Andrade Jr.}}}]{Araujo03}%
  \BibitemOpen
  \bibfield  {author} {\bibinfo {author} {\bibfnamefont {A.~D.}\ \bibnamefont
  {Ara\'ujo}}, \bibinfo {author} {\bibfnamefont {A.~A.}\ \bibnamefont
  {Moreira}}, \bibinfo {author} {\bibfnamefont {R.~N.}\ \bibnamefont {{Costa
  Filho}}}, \ and\ \bibinfo {author} {\bibfnamefont {J.~S.}\ \bibnamefont
  {{Andrade Jr.}}},\ }\href {\doibase 10.1103/PhysRevE.67.027102} {\bibfield
  {journal} {\bibinfo  {journal} {Phys. Rev. E}\ }\textbf {\bibinfo {volume}
  {67}},\ \bibinfo {pages} {027102} (\bibinfo {year} {2003})}\BibitemShut
  {NoStop}%
\bibitem [{\citenamefont {Du}\ \emph {et~al.}(1996)\citenamefont {Du},
  \citenamefont {Satik},\ and\ \citenamefont {Yortsos}}]{Du04}%
  \BibitemOpen
  \bibfield  {author} {\bibinfo {author} {\bibfnamefont {C.}~\bibnamefont
  {Du}}, \bibinfo {author} {\bibfnamefont {C.}~\bibnamefont {Satik}}, \ and\
  \bibinfo {author} {\bibfnamefont {Y.~C.}\ \bibnamefont {Yortsos}},\ }\href
  {\doibase 10.1002/aic.690420831} {\bibfield  {journal} {\bibinfo  {journal}
  {AIChE Journal}\ }\textbf {\bibinfo {volume} {42}},\ \bibinfo {pages} {2392}
  (\bibinfo {year} {1996})}\BibitemShut {NoStop}%
\bibitem [{\citenamefont {Lauritsen}\ \emph {et~al.}(1993)\citenamefont
  {Lauritsen}, \citenamefont {Sahimi},\ and\ \citenamefont
  {Herrmann}}]{Lauritsen93}%
  \BibitemOpen
  \bibfield  {author} {\bibinfo {author} {\bibfnamefont {K.~B.}\ \bibnamefont
  {Lauritsen}}, \bibinfo {author} {\bibfnamefont {M.}~\bibnamefont {Sahimi}}, \
  and\ \bibinfo {author} {\bibfnamefont {H.~J.}\ \bibnamefont {Herrmann}},\
  }\href {\doibase 10.1103/PhysRevE.48.1272} {\bibfield  {journal} {\bibinfo
  {journal} {Phys. Rev. E}\ }\textbf {\bibinfo {volume} {48}},\ \bibinfo
  {pages} {1272} (\bibinfo {year} {1993})}\BibitemShut {NoStop}%
\bibitem [{\citenamefont {Mamede}\ \emph {et~al.}(2012)\citenamefont {Mamede},
  \citenamefont {Ara\'ujo}, \citenamefont {Schneider}, \citenamefont {{de
  Ara\'ujo}},\ and\ \citenamefont {Herrmann}}]{Mamede12}%
  \BibitemOpen
  \bibfield  {author} {\bibinfo {author} {\bibfnamefont {G.~L.}\ \bibnamefont
  {Mamede}}, \bibinfo {author} {\bibfnamefont {N.~A.~M.}\ \bibnamefont
  {Ara\'ujo}}, \bibinfo {author} {\bibfnamefont {C.~M.}\ \bibnamefont
  {Schneider}}, \bibinfo {author} {\bibfnamefont {J.~C.}\ \bibnamefont {{de
  Ara\'ujo}}}, \ and\ \bibinfo {author} {\bibfnamefont {H.~J.}\ \bibnamefont
  {Herrmann}},\ }\href {\doibase 10.1073/pnas.1200398109} {\bibfield  {journal}
  {\bibinfo  {journal} {Proc. Natl. Acad. Sci. USA}\ }\textbf {\bibinfo
  {volume} {109}},\ \bibinfo {pages} {7191} (\bibinfo {year}
  {2012})}\BibitemShut {NoStop}%
\bibitem [{\citenamefont {Schrenk}\ \emph
  {et~al.}(2012{\natexlab{b}})\citenamefont {Schrenk}, \citenamefont
  {Ara\'ujo},\ and\ \citenamefont {Herrmann}}]{Schrenk12b}%
  \BibitemOpen
  \bibfield  {author} {\bibinfo {author} {\bibfnamefont {K.~J.}\ \bibnamefont
  {Schrenk}}, \bibinfo {author} {\bibfnamefont {N.~A.~M.}\ \bibnamefont
  {Ara\'ujo}}, \ and\ \bibinfo {author} {\bibfnamefont {H.~J.}\ \bibnamefont
  {Herrmann}},\ }\href {\doibase 10.1038/srep00751} {\bibfield  {journal}
  {\bibinfo  {journal} {Sci. Rep.}\ }\textbf {\bibinfo {volume} {2}},\ \bibinfo
  {pages} {751} (\bibinfo {year} {2012}{\natexlab{b}})}\BibitemShut {NoStop}%
\end{thebibliography}%

\end{document}